# Technological Aspects: High Voltage

*D.C. Faircloth*
Rutherford Appleton Laboratory, Chilton, UK

**Abstract**
This paper covers the theory and technological aspects of high-voltage design for ion sources. Electric field strengths are critical to understanding high-voltage breakdown. The equations governing electric fields and the techniques to solve them are discussed. The fundamental physics of high-voltage breakdown and electrical discharges are outlined. Different types of electrical discharges are catalogued and their behaviour in environments ranging from air to vacuum are detailed. The importance of surfaces is discussed. The principles of designing electrodes and insulators are introduced. The use of high-voltage platforms and their relation to system design are discussed. The use of commercially available high-voltage technology such as connectors, feedthroughs and cables are considered. Different power supply technologies and their procurement are briefly outlined. High-voltage safety, electric shocks and system design rules are covered.

## 1 Introduction

### 1.1 Uses

High voltages (HV) are a fundamental component of ion sources, they are used for:

– extracting beams from plasmas (up to 50 kV),

– accelerating beams (up to 28 000 kV),

– initiating discharges and pre-ionizing gases (up to 20 kV),

– focusing and deflecting beams (up to 50 kV),

– suppressing unwanted particles (up to 5 kV).

### 1.2 Challenges

Ion sources are particularly challenging for HV design because they often operate at high temperatures in magnetic fields with large numbers of free charge carriers and stray beams. In negative ion sources, caesium often coats surfaces and encourages breakdowns. To keep efficiencies high and to minimize emittance and beam size, ion sources are designed to be compact. Unfortunately, this only makes the high-voltage aspect of the design even more difficult. There are very rarely any experimental data available for the specific combination of conditions present to aid design calculation. Therefore new source designs often have to undergo several prototype iterations. Modelling can help point to problems, but test-stands and a 'suck it and see' approach are essential because problems often emerge only after extended periods of operation.

High voltages are an obvious electrocution safety hazard and must be treated with great care. Safety systems must be put in place to prevent accidental electric shock. Also, ion sources often run with explosive gases such as hydrogen, and high-voltage sparks are an ignition source. Compliance with European ATEX directives [1] is essential.

## 1.3 Aims

The main aim of high-voltage design for ion sources is as follows: to produce reliable breakdown where it is wanted and to prevent breakdown where it is unwanted.

Reliable breakdown must occur in the discharge volume and any pre-ignition system. Breakdown must be prevented (or its occurrence minimized to acceptable levels) in the extraction, acceleration, focusing, deflection and suppression systems. Particular care must be given to the design of the extraction system because it is always right next to the discharge volume.

## 1.4 Factors affecting HV breakdown

High-voltage breakdown occurs when the electric field in a system becomes high enough to cause electron avalanches (see section 3.4). Electric field is the potential gradient, or the rate at which the voltage changes per unit length. It has many names: electric field strength, electric field intensity, stress, or just $E$. It is expressed in units of $V\,m^{-1}$, $kV\,m^{-1}$, $kV\,mm^{-1}$, $kV\,cm^{-1}$ or $MV\,m^{-1}$.

In general, high-voltage breakdown is most likely to occur where the electric field is the highest, but this depends on various factors. Different insulating materials under different conditions will break down at different electric field strengths. The electric field where a material breaks down is called the material's dielectric strength and is expressed in the same units as electric field.

Different gases have different dielectric strengths. Increasing a gas's temperature will decrease its dielectric strength, as will decreasing its pressure. However, at very low pressures the dielectric strength sharply increases again, as we enter the realm of vacuum insulation (see section 3.7 on the Paschen curve).

Solid insulators generally have a much greater dielectric strength than gaseous ones. Gaseous and liquid insulators generally recover after breakdown, whereas breakdown of solid insulators usually causes irreversible damage.

The geometry of the electrodes plays a critical role in the type of discharge produced and the overall breakdown voltage. Electrode surface roughness can also reduce the voltage at which a system breaks down by causing localized field enhancements on a microscopic level. This is why high-voltage electrodes are often highly polished.

Another very important factor determining the voltage at which a system will break down is the presence of insulating surfaces. Surfaces are inevitable in high-voltage systems because the electrodes must be supported somehow. Solid insulators may have a very high dielectric strength in their bulk volume (i.e., their resistance to being punctured by the high field.) However, their surfaces can be very weak. Once initiated, discharges can propagate along insulator surfaces in very low field strengths. This often causes permanent damage to the insulator surface.

The point where the insulator surface meets the electrode can lead to localized field enhancements that can initiate discharges, dramatically lowering the breakdown voltage (see section 4.3.2 on triple junction effects).

Insulators and insulator surfaces can charge up, affecting the electric field. This can either increase or decrease the voltage at which the system breaks down depending on polarity. The actual mechanism of high-voltage breakdown is statistical by nature. No two sparks are ever the same. This combined with the complexities of charging mean that it is impossible to predict the exact voltage at which a system will break down.

## 2 Calculating electric fields

### 2.1 Introduction

Knowing the electric field strengths present in a high-voltage system is essential to predicting its performance. Electric fields can be calculated using equations, measured in experimental analogues, or solved with numerical techniques.

### 2.2 Maxwell's equations

Maxwell's equations are a set of four equations that completely describe the relationships between electric and magnetic fields. They define not only the shape and strength of the electromagnetic field, but also how it varies with time. The equations can be expressed in different ways, but the most compact form is in differential notation:

$$\nabla \times \mathbf{E} = -\frac{\partial \mathbf{B}}{\partial t} \tag{1}$$

$$\nabla \times \mathbf{H} = \mathbf{J} + \frac{\partial \mathbf{D}}{\partial t} \tag{2}$$

$$\nabla \cdot \mathbf{B} = 0 \tag{3}$$

$$\nabla \cdot \mathbf{D} = \rho \tag{4}$$

where

**E** is the electric field strength,

**B** is the magnetic flux density,

**H** is the magnetic field strength,

**J** is the current density,

**D** is the electric flux density,

$\rho$ is the charge density.

Equation (1) describes how a time-varying magnetic field generates an electric field. It explains how electrical generators work. It is the differential form of Faraday's law of induction.

Equation (2) describes how electrical currents produce magnetic fields. It explains how electromagnets and motors work.

Equation (3) describes the shape of the magnetic field. It implies that magnetic monopoles do not exist.

Equation (4) describes the shape of the electric field in the presence of electrical charges. Both point and diffuse charges can exist. It is the differential form of Gauss's law.

### 2.3 Poisson and Laplace equations

#### 2.3.1 *Derivation*

If no magnetic fields are present Eq. (1) becomes

$$\nabla \times \mathbf{E} = 0 \tag{5}$$

The definition of electric field is the rate at which the scalar potential $\phi$ varies with distance:
$$\mathbf{E} = -\operatorname{grad} \phi \tag{6}$$

The electric field and electric displacement field (electric flux density) are related by the electrical permittivity $\varepsilon$:
$$\mathbf{D} = \varepsilon \mathbf{E} \tag{7}$$

Substituting Eq. (6) into Eq. (7) gives
$$\mathbf{D} = -\varepsilon \operatorname{grad} \phi \tag{8}$$

Substituting Eq. (8) into Eq. (4) gives Poisson's equation
$$\nabla^2 \phi = \frac{\rho}{\varepsilon} \tag{9}$$

which can also be written as
$$\frac{\partial^2 \phi}{\partial x^2} + \frac{\partial^2 \phi}{\partial y^2} + \frac{\partial^2 \phi}{\partial z^2} = \frac{\rho}{\varepsilon} \tag{10}$$

If no charges are present, Eq. (10) becomes Laplace's equation:
$$\frac{\partial^2 \phi}{\partial x^2} + \frac{\partial^2 \phi}{\partial y^2} + \frac{\partial^2 \phi}{\partial z^2} = 0 \tag{11}$$

## 2.4 Solving Laplace's equation

### 2.4.1 Infinite parallel-plate capacitor

Laplace's equation can be used to calculate the electric fields present in different geometries. Consider an infinite parallel-plate capacitor, as shown in Fig. 1. One plate is grounded, and the other is held at a potential $V$. The plates are a distance $d$ apart.

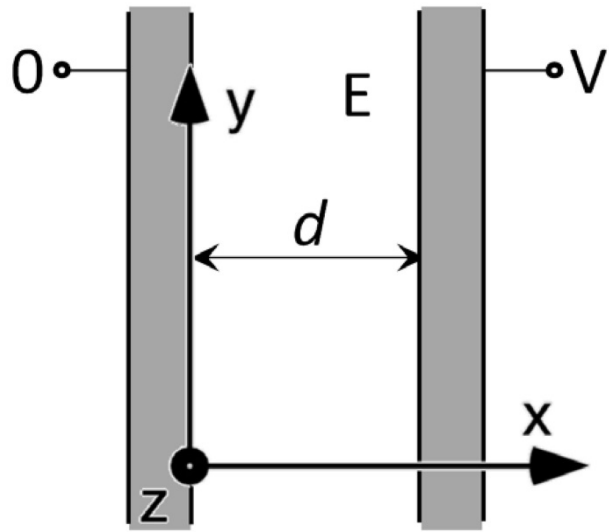

**Fig. 1**: Infinite parallel-plate capacitor

If the axis are defined as shown in Fig. 1, then the potential $\phi$ between the plates does not vary in the $y$ or $z$ directions:

$$\frac{\partial \phi}{\partial y} = \frac{\partial \phi}{\partial z} = 0 \tag{12}$$

Therefore the second derivative is also zero:

$$\frac{\partial^2 \phi}{\partial y^2} = \frac{\partial^2 \phi}{\partial z^2} = 0 \tag{13}$$

Substituting into the Laplace equation (11) gives

$$\frac{\partial^2 \phi}{\partial x^2} = 0 \tag{14}$$

Integrating this gives

$$\frac{\partial \phi}{\partial x} = c_1 \tag{15}$$

and integrating again gives

$$\phi(x) = c_1 x + c_2 \tag{16}$$

At $x = 0$, $\phi = 0$, and at $x = d$, $\phi = V$, so

$$c_1 = \frac{V}{d} \tag{17}$$

and

$$c_2 = 0 \tag{18}$$

So Eq. (16) becomes

$$\phi(x) = \frac{V}{d} x \tag{19}$$

From Eq. (6) we get

$$\mathbf{E} = -\frac{V}{d} \tag{20}$$

or

$$|\mathbf{E}| = \frac{V}{d} \tag{21}$$

### 2.4.2 Infinite coaxial line

Consider and infinite coaxial line as shown in Fig. 2. The inner cylindrical conductor is held at a potential $\phi_1$ and the outer cylindrical conductor is held at a potential $\phi_2$. The radii of the inner and outer conductors are $r_1$ and $r_2$, respectively.

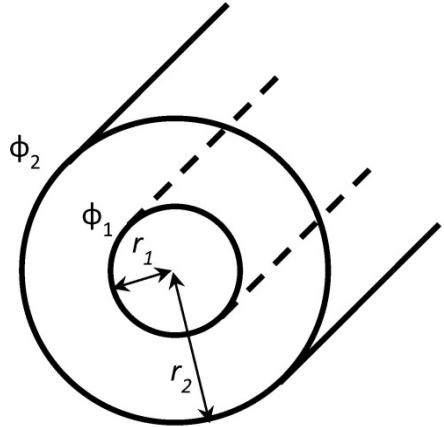

**Fig. 2:** Infinite coaxial line

The Laplace equation (11) can be expressed in terms of cylindrical symmetric coordinates:

$$\frac{1}{r}\cdot\frac{\partial}{\partial r}\left(r\frac{\partial \phi}{\partial r}\right) + \frac{1}{r^2}\frac{\partial^2 \phi}{\partial \varphi^2} + \frac{\partial^2 \phi}{\partial z^2} = 0 \tag{22}$$

If the axes in Fig. 2 are defined so that the z-axis runs along the centre of the coaxial line, then the voltage $\phi$ does not vary with angle $\varphi$ or distance $z$. So

$$\frac{\partial^2 \phi}{\partial \varphi^2} = \frac{\partial^2 \phi}{\partial z^2} = 0 \tag{23}$$

Substituting into the cylindrical systematic form of Laplace's equation (22) gives

$$\frac{1}{r}\cdot\frac{\partial}{\partial r}\left(r\frac{\partial \phi}{\partial r}\right) = 0 \tag{24}$$

By integration and substitution we get

$$\phi(r) = \phi_1 + \frac{\phi_1 - \phi_2}{\ln\left(\dfrac{r_1}{r_2}\right)}\left(\ln r - \ln r_1\right) \tag{25}$$

From Eq. (6),

$$E(r) = \frac{\phi_1 - \phi_2}{r\ln\left(\dfrac{r_2}{r_1}\right)} \tag{26}$$

The field is greatest on the surface of the inner conductor where $r = r_1$:

$$E_{max} = \frac{\phi_1 - \phi_2}{r_1 \ln\left(\frac{r_2}{r_1}\right)} \tag{27}$$

## 2.5 Electric fields measured in experimental analogues

### 2.5.1 Introduction

Algebraically solving the electric field distribution is relatively easy for the idealized geometries shown in the previous section. It is possible to solve the fields in real-world geometries as long as the electrode surfaces are made up of basic shapes, although as the electrode geometries become more complex the equations rapidly become very difficult to solve. When the electrode surface geometry cannot accurately be described by an equation, then algebraic solution of the fields is no longer possible.

Although rarely used these days, it is possible to set up an experimental analogue to directly measure electric potentials using a model of the electrode geometry.

### 2.5.2 Teledeltos paper

Teledeltos paper is an electrically conductive paper. It is formed by coating carbon on one side of a sheet of paper. This provides a sheet of uniform resistance, with isotropic resistivity in each direction.

Teledeltos paper can be used to create a two-dimensional model of any scalar field, such as the electric field. Conductive paint is used to paint the shape of the electrodes on the paper. Power supplies are used to apply voltages to the painted-on electrodes. A probe connected to a voltmeter is then used to measure the voltage at various points between the painted-on electrodes. In this way a map of the scalar electric potential $\phi$ can be constructed. Then equipotential lines can be drawn and electric field strengths calculated.

The resistances involved are in the kilohm range. This is low enough that the Teledeltos paper may be used with safe low voltages, yet high enough that the currents remain low, avoiding problems with contact resistance.

### 2.5.3 Electrolytic tank

A similar technique to the Teledeltos paper method involves the use of a tank of conductive fluid [2]. Scaled or actual size copies of the electrodes are placed in a tank of fluid, which could be a solution of $NiSO_4$, $CuSO_4$ or just pure distilled water. Voltages are applied to the electrodes and a probe is used to measure the voltage at various positions to generate a map of scalar electric potential in the system.

Using a fluid instead of a conductive surface has two main advantages: three-dimensional field maps can be generated; and the actual electrodes can be tested rather than conductive paint drawings.

In reality it is quite difficult to accurately measure three-dimensional field maps, so electrolytic tanks were often used with a layer of fluid to generate two-dimensional field maps instead. By tilting the tank, it is possible to generate two-dimensional field maps at different 'slices' through the electrode geometry under test.

Figure 3 shows a sophisticated electrolytic tank system capable of automatically plotting equipotential lines. A simple electrode geometry consisting of two circular electrodes is being tested. The electrodes sit in a tank of electrolytic fluid and a voltage is applied between them. The voltage measurement probe is attached to the lid of the tank and can be moved in the *x* and *y* directions using two motor drives. When the lid is closed, the probe is immersed in the electrolytic solution and, with the help of control electronics, starts to follow an equipotential line. A pen is mounted on the probe, which plots the trajectory of the probe on a sheet of paper mounted on the lid, thus drawing the equipotential line. This process is repeated for different probe voltages until an adequate map of the field between the electrodes is obtained.

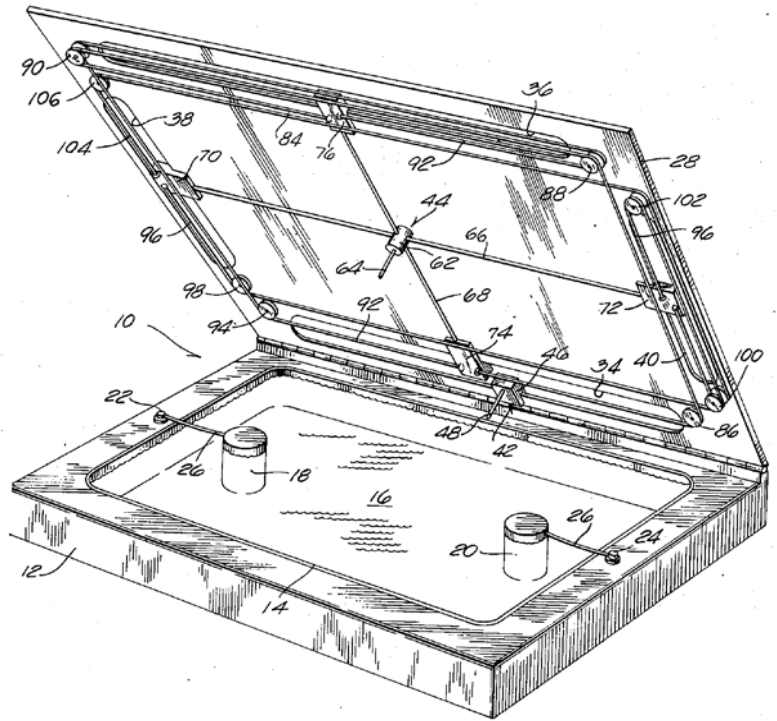

**Fig. 3**: An automated electrolytic tank measuring the equipotentials between two circular electrodes (from United States Patent No. 3764900).

## 2.6 Solving electric field distributions with numerical techniques

### 2.6.1 Introduction

Thankfully, plotting fields with experimental analogues as described in the previous section is now a thing of the past. Today, experimental analogues are only used in student laboratories as demonstrations. The increased availability of computing power from the 1970s onwards has made solving electric fields using numerical methods commonplace.

### 2.6.2 Finite elements

The finite element technique involves breaking the electrode geometry and the space between the electrodes into discrete regions called elements. This process is called meshing because the resultant pattern resembles the mesh in a fisherman's net, or the delicate mesh pattern in a net curtain. A typical two-dimensional finite element mesh is shown in Fig. 4.

The corners of each element are called nodes. Using the discrete form of Poisson's equation, it is possible to define the potential of each node in terms of its neighbour's potentials and their distance away (the length of the element's side). In this way a set of simultaneous equations defining the potential of every node can be defined. It is then just a case of solving all the simultaneous equations – a problem ideally suited to computers.

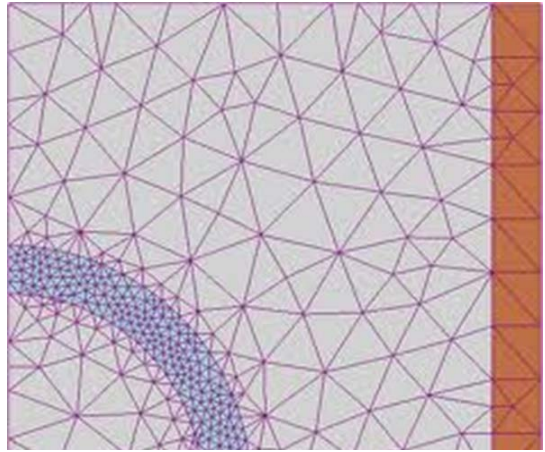

**Fig. 4**: Finite elements in a two-dimensional mesh

### 2.6.3   Element shape and size

The example shown in Fig. 4 is two-dimensional and uses triangular elements, but it is equally possible to use square or rectangular elements. Three-dimensional problems can also be solved using tetrahedral or brick-shaped elements.

The accuracy of the field distribution obtained depends on the size of the elements: smaller elements and more dense elements will yield higher-accuracy results. However, smaller elements means more elements. More elements means more simultaneous equations to solve. It takes longer to solve more simultaneous equations, so more computing resources are required. Therefore, when meshing a geometry, it is always important to be as efficient with element size as possible: do not make the mesh high density everywhere, only where it is needed. This is most important in three dimensions, where very large problems can take hours to solve.

### 2.6.4   Boundaries

To solve simultaneous equations, there must be some known values. In electric field problems, these are nodes that lie on the conducting electrode surfaces. Electrodes that have known voltages applied to them will have their nodes set to that specific voltage. Electrodes that are left to float in potential can have a condition imposed to make all the nodes have an equal value.

An important technique used to reduce the total number of elements in a model is the use of symmetry planes. If the electrode geometry has reflectional symmetry, then only half the geometry needs to be modelled. The equipotential field lines pass through the line of reflectional symmetry at right angles. This means there is no electric field component normal to the line of symmetry. In the model, the line of symmetry becomes a boundary where the electric field normal to the boundary must be zero. This imposes a condition on the nodes on that boundary.

Rotational symmetries can also be exploited to reduce the number of elements. This is especially true for three-dimensional geometries with an axis of rotational symmetry. The geometry can be represented as a two-dimensional plane, which will be much quicker to solve and give an identical result to solving the full three-dimensional geometry.

It may also be possible to reduce a three-dimensional geometry to a two-dimensional geometry by taking a slice through it. However, unless the objects being modelled are very long in the direction orthogonal to the slice, this will only be an approximation to the true field.

### 2.6.5  Solution methods

There are many ways to solve the simultaneous equations produced by the finite element technique. Here just are a few of the more commonly used techniques:

*Direct methods:*
- Gaussian elimination
- LU decomposition method

*Iterative methods:*
- Mesh relaxation methods
  - Jacobi
  - Gauss–Seidel
  - Successive over-relaxation (SOR) method
  - Alternating directions implicit (ADI) method
- Matrix methods
  - Thomas tridiagonal form
  - Sparse matrix methods
  - Conjugate gradient (CG) methods
  - Multi-grid (MG) method

It is perfectly possible to write your own solver, and this is the subject of many undergraduate and Masters degree projects; however, you will be reinventing the wheel. There are plenty of commercial and open-source codes available to do this written in every language for every platform.

### 2.6.6  Professional modelling software

There are many advantages to using existing codes other than time saved in not having to develop your own code. Efficiently meshing a geometry, especially a three-dimensional one, is a difficult mathematical problem in itself. Even the expensive commercially written codes sometimes have problems meshing geometries containing a lot of fine details.

There are many codes available. A particularly good free two-dimensional code is Poisson Superfish, developed by the code group at Los Alamos National Laboratory in the USA [3]. Here is a non-exhaustive list of some commercially available codes:

- ANSYS
- Cobham Opera
- COMSOL
- CST Studio Suite
- SimIon
- INP

The commercially available codes can be expensive, especially the three-dimensional ones that are capable of solving magnetic and other transient problems. The advantages of professional codes are:

- Quick and easy geometry importing directly from CAD files.
- Advanced meshing tools.
- Multicore solvers.
- Proven benchmarked code.
- User support helpdesks and large user communities.

# 3 Electrical discharges

## 3.1 Introduction

There are many different types of electrical discharges. Some are found in nature, such as lightning, aurora and Saint Elmo's fire. This section will cover the different types of electrical discharges found in engineering. They can be classified in different ways: Are they localized or do they completely bridge the gap between the electrodes? Are they transient or continuous? Are they wanted or unwanted?

'Wanted' electrical discharges are used to create the plasma in ion sources, and these can be glow or arc discharges (see sections 3.13 and 3.14). These types of discharges must be reliably and repeatably produced. Starting a discharge is referred to as striking the discharge.

'Unwanted' discharges in high-voltage systems can lead to catastrophic failure, but not always: it depends if the discharge remains localized in a medium that self-heals. A good example is a corona discharge from a sharp point in air (see section 3.11). If the system geometry is large enough, then the corona can fizz away without much problem. The corona will cause a few extra microamps to be drawn from the high-voltage power supply, create electromagnetic noise, and a small amount of ozone, but it will not lead to an overall failure of the high-voltage system.

However, if the localized discharge is in a non-self-healing medium, then problems can occur. A good example is a void (or hole) in solid insulation (see section 3.10). Initially the discharge remains localized, just causing electromagnetic noise, but slowly it will damage the insulator material and potentially lead to a catastrophic failure.

An 'unwanted' discharge that completely bridges the gap between the electrodes is referred to as a 'breakdown' or a 'flashover' and is often visible as a spark accompanied by an audible crack. If a solid insulating medium is between the electrodes and the discharge breaks through it, the insulator is said to have been 'punctured' or 'ruptured'. The weakest point of an insulator is its surface (see section 3.9.1). A discharge that propagates along the surface is referred to as 'tracking' because it often leaves carbonized tree- or fern-like tracks on the insulator surface. A similar term in solid insulation is 'treeing' (see Fig. 5). The paths left by high-voltage discharges can be quite beautiful.

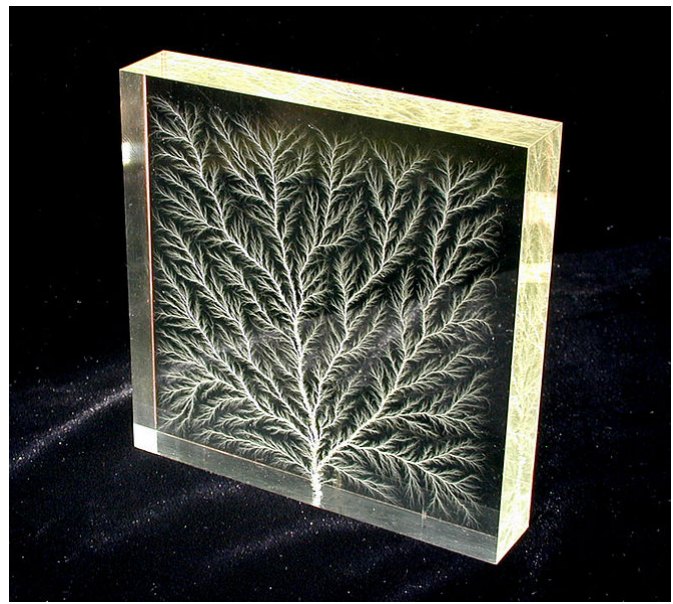

**Fig. 5:** Treeing in a block of Plexiglas. © Bert Hickman

Electrical discharges are plasmas. Plasmas are chemically reactive, so they can cause chemical damage to insulators. Plasmas contain fast-moving ions that can damage electrodes by sputtering. This must be considered when designing the plasma chamber for ion sources. Discharges also produce a significant amount of ultraviolet (UV) light that can degrade insulator materials.

The route to high-voltage breakdown depends on the degree of field uniformity in the system. Discharges can be initiated in localized regions of high field strength, then continue to propagate into regions where the field is not strong enough to start a discharge, but it is strong enough to allow the discharge to continue to propagate. A well-designed high-voltage system removes regions of high field by rounding electrodes and screening sharp points (see section 4.2).

Whether the discharge is transient or continuous can also depend on how the discharge gets its power (see section 3.12).

## 3.2    Ionization

### 3.2.1    *Background ionization*

Electrical discharges are fundamentally about ionizing materials (gases, liquids or solids) to produce charge carriers (electrons and ions). There is always a low level of background ionization occurring due to background radiation. Background radiation comes mainly from natural sources such as radon gas and cosmic rays but also includes a small contribution from manmade nuclear tests and accidents.

### 3.2.2    *Electron impact ionization*

The most important ionization mechanism in electrical discharges is electron impact ionization. Free electrons impact on the outer electron orbitals of atoms and molecules all the time. If the free electrons have enough energy, they can knock the outer electron out of its orbital, leaving a positive ion and another free electron. The free electrons get their energy from externally applied fields. The field applied to create a discharge is usually the electric field; however, a magnetic field can also be applied in the case of inductively coupled discharges.

Inductively coupled discharges can only be produced with a time-varying magnetic field. The free electrons are actually accelerated by the spatially varying, non-conservative electric field produced by the time-varying magnetic field. This is the Maxwell–Faraday law of electromagnetic induction shown in Eq. (1).

Electric-field-driven discharges can also be produced by a.c. electric fields, but they extinguish and reignite each time the electric field reverses polarity.

### 3.2.3    *Photo-ionization*

Photons with enough energy can also cause ionization. These are usually photons in the UV part of the spectrum. The discharge itself can create UV light when excited atoms relax or ions recombine with electrons. This plays an important role in helping the discharge to propagate.

An excited (or metastable) state is where the electrons are still attached to the atom; they occupy higher 'excited' orbitals. An exited atom can go on to be completely ionized by absorbing more energy, either from another photon or by being hit by a free electron.

If the excited atom does not receive any more energy, it will relax back into a lower state by emitting a photon, which could go on to excite the electrons in another atom.

## 3.3    Discharge current/voltage landscape

There are many names for the basic phenomenon that comprise electrical discharges, such as avalanches, streamers, corona, leaders, glows and arcs. These will all be defined in the rest of this section. They cover a huge range of different currents. To provide a conceptual landscape in which these discharges exist, it is useful to consider a simple set-up with two electrodes shown in Fig. 6(a). A d.c. voltage can be applied between the electrodes and the resulting current measured.

The general current–voltage characteristic of such a discharge is summarized in Fig. 6(b). The exact shape of the curve depends on the type of gas, pressure, electrode geometry, electrode temperatures, electrode materials and any magnetic fields present.

At low voltages, the current between two electrodes is extremely small, but it slowly increases as the voltage between the electrodes increases (as shown in the bottom left corner of the graph in Fig. 6(b)). This tiny current comes from charge carriers produced by background ionization. They are swept out of the gap by the electric field between the electrodes that is created by the applied voltage. There are only enough charge carriers produced by background radiation for a few nanoamps of current, so the current quickly saturates. The voltage can then be increased with no increase in current. The ions and electrons are pulled towards the electrodes through the gas molecules interacting with them as they go.

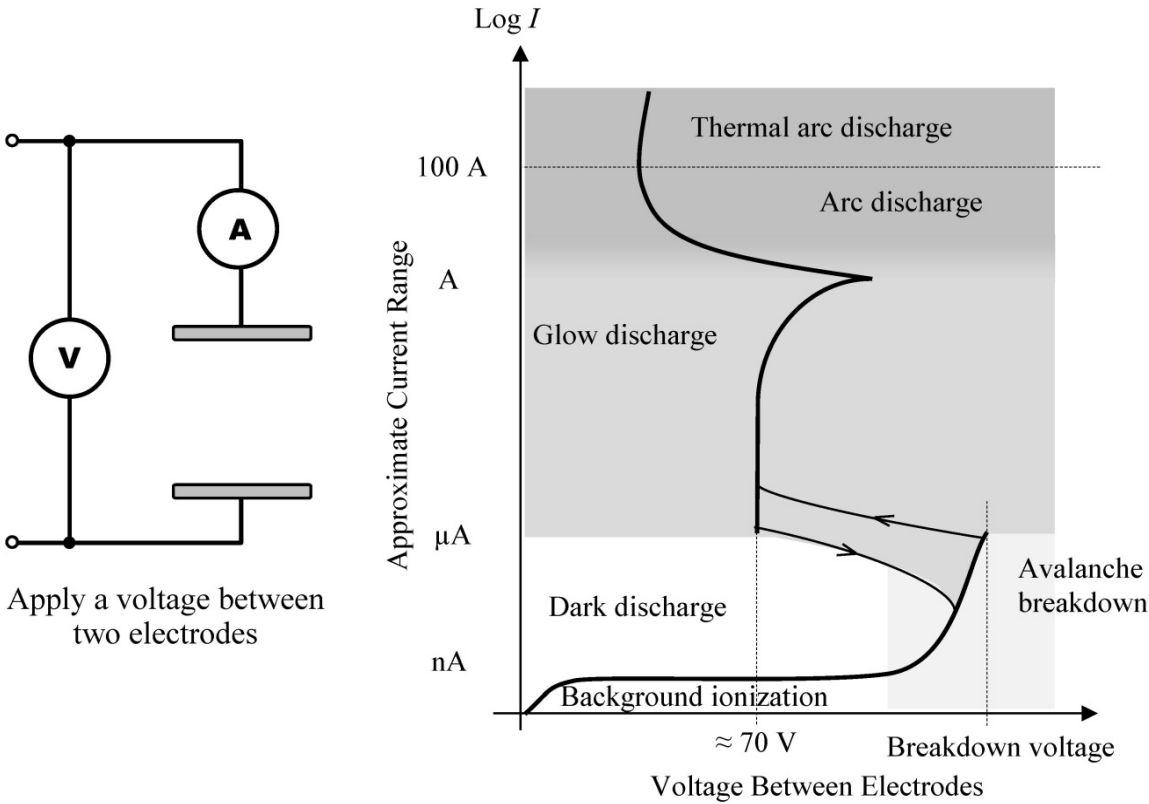

**Fig. 6:** The current–voltage characteristic of a typical electrical discharge between two electrodes

### 3.4 Avalanche breakdown

#### 3.4.1 Introduction

The voltage between the electrodes shown in Fig. 6(a) is increased until the applied electric field is high enough to accelerate the electrons to the ionization energy of the gas. At this point the current rapidly increases, as shown in the bottom right corner of the graph in Fig. 6(b). The electrons ionize the neutral atoms and molecules, producing more electrons. These additional electrons are accelerated to ionize even more atoms, producing even more free electrons in an avalanche build-up process, as shown in Fig. 7.

This is the moment of inception of a high-voltage breakdown, and if the conditions are right, it can cause complete flashover of the electrodes.

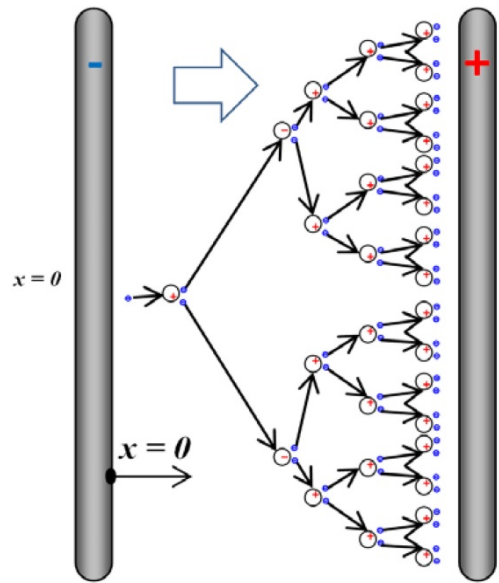

**Fig. 7:** The avalanche breakdown mechanism in electrical discharges

#### 3.4.2 Townsend's primary ionization coefficient, α

The avalanche process was first mathematically described by John Townsend in 1897. Consider an avalanche discharge between two electrodes. The position between the electrodes in Fig. 7 is given by the variable $x$. The number of additional free electrons $dn_x$ produced in a distance $dx$ depends on the number of electrons at that point $n_x$ and the primary ionization coefficient $\alpha$:

$$dn_x = n_x \alpha \, dx \tag{28}$$

The primary ionization coefficient $\alpha$ is the number of additional electrons produced per unit length. By integration and the fact that $n_x = n_0$ at $x = 0$, this gives

$$n_x = n_0 \, e^{\alpha x} \tag{29}$$

The number of free electrons (and ions) increases exponentially in the avalanche.

## 3.5 Streamers

### 3.5.1 Electric field in an avalanche

Figure 8(a) shows the distribution of particles in a single avalanche. The discharge has a negative head, comprising the avalanche of free electrons, and a positive tail, comprising the positive ions left behind after the ionization avalanche has passed. The ions are at least 1800 times heavier than the electrons so they take much longer to accelerate than the electrons. Compared to the fast free electrons, the ions only move a tiny distance from where they were born.

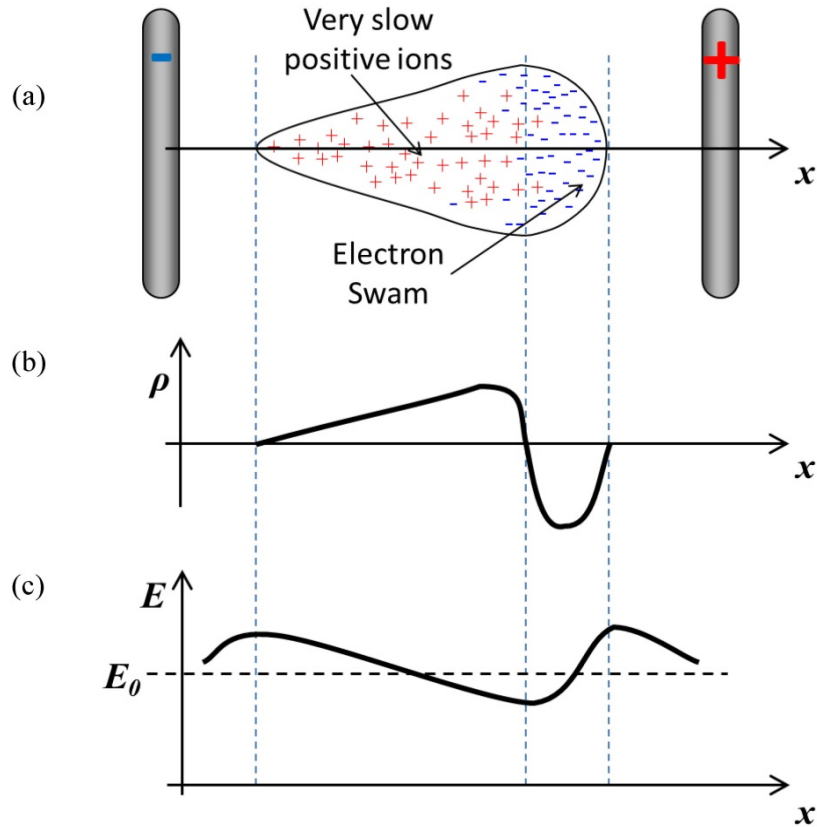

**Fig. 8**: Distribution of (a) particles, (b) charge density and (c) electric field distribution in a single avalanche discharge.

The difference in mobility of the ions and electrons creates the charge distribution shown in Fig. 8(b), which in turn produces the electric field distribution shown in Fig. 8(c). If the discharge was not present, the electric field between the electrodes would be $E_0$. Even though the total amount of positive and negative particles is approximately equal, the distribution of charges in the avalanche causes an increase in the field in front of and behind the discharge. This is caused by the negative electron avalanche being closer to the positive electrode and the positive ions being closer to the negative electrode.

### 3.5.2 Streamer formation

The avalanche emits photons that can ionize nearby atoms creating free electrons as described in section 3.2.3. Any free electrons created in the higher-field region in front of or behind the initial avalanche will go on to produce additional avalanches, as shown in Fig. 9.

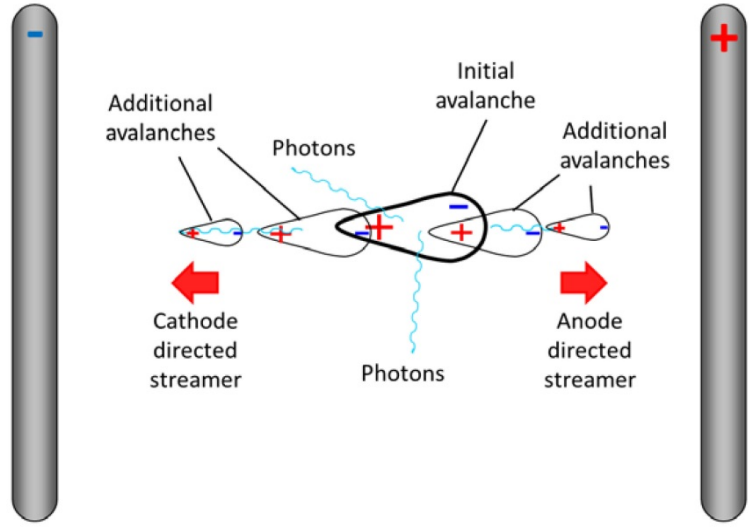

**Fig. 9**: Streamer formation made up of a series of avalanches propagating in both directions from the initial avalanche.

Each additional avalanche creates more photons and further enhances the electric field. In this way a chain of avalanches propagates from the head and tail of the initial avalanche. The name for this chain of avalanches is a streamer. Streamers propagate in both directions, as shown in Fig. 9. Eventually the chain of avalanches bridges the gap between the electrodes, creating an ionized conductive channel. This is the moment the gap actually flashes over. As much current flows from one electrode to the other as the power supply can deliver.

### 3.5.3  Townsend's secondary ionization coefficient, γ

Townsend also considered secondary ionization processes. In addition to photoionization and metastable effects, ion impact ionization should also be considered. This is where the positive ions impact on neutral atoms or the cathode electrode surface. Positive ions impacting on the cathode significantly enhance the growth of the discharge at the cathode.

The secondary ionization coefficient $\gamma$ is the number of secondary electrons produced per electron in the primary avalanche. It includes all secondary processes:

$$\gamma = \gamma_p + \gamma_m + \gamma_{ion} \tag{30}$$

Take a pair of electrodes a distance $d$ apart and assume all electrons that leave the cathode and are produced by primary and secondary processes reach the anode. Then using Eq. (29) the total current in the discharge can be shown to be

$$I = \frac{I_0 e^{\alpha d}}{1 - \lambda(e^{\alpha d} - 1)} \tag{31}$$

When breakdown occurs, the current increases very rapidly. This allows a criterion for breakdown to be obtained from Eq. (31):

$$\gamma(e^{\alpha d} - 1) = 1 \tag{32}$$

The term $e^{\alpha d}$ is usually very large, so Eq. (32) becomes

$$\gamma e^{\alpha d} = 1 \tag{33}$$

This is the Townsend criterion for breakdown and gives the conditions required for a self-sustaining discharge. Eq. (32) can be used when modelling breakdowns.

## 3.6 Leaders

Leaders only occur in very long breakdowns (more than a metre) so they only happen when using extra-high voltages (EHV). Pre-injectors and tandems use EHV d.c. to accelerate beams after extraction from the ion source.

A leader is a hot plasma channel that starts from either the positive or negative electrode. The current from many streamers forms a highly ionized region at the electrode surface, which then propagates into the gap. The highly conductive leader channel is effectively at the same potential as the electrode. Leader growth is significantly slower than the fast streamers, and it propagates by creating more streamers at its head.

Leader formation is a vital component of lighting propagation, but its presence in high-voltage engineering is a rare and unwanted phenomenon.

## 3.7 Paschen curve

The breakdown voltage of a gas between two flat electrodes depends only on the electron mean free path and the distance between the electrodes. The electron mean free path is the average distance the electrons travel before hitting atoms, so it is the key factor defining the growth of an avalanche discharge. It is directly related to pressure. Fig. 10 shows how the breakdown voltage of hydrogen varies with the product of pressure $p$ and distance $d$ between electrodes. This was first stated in 1889 by Friedrich Paschen [4].

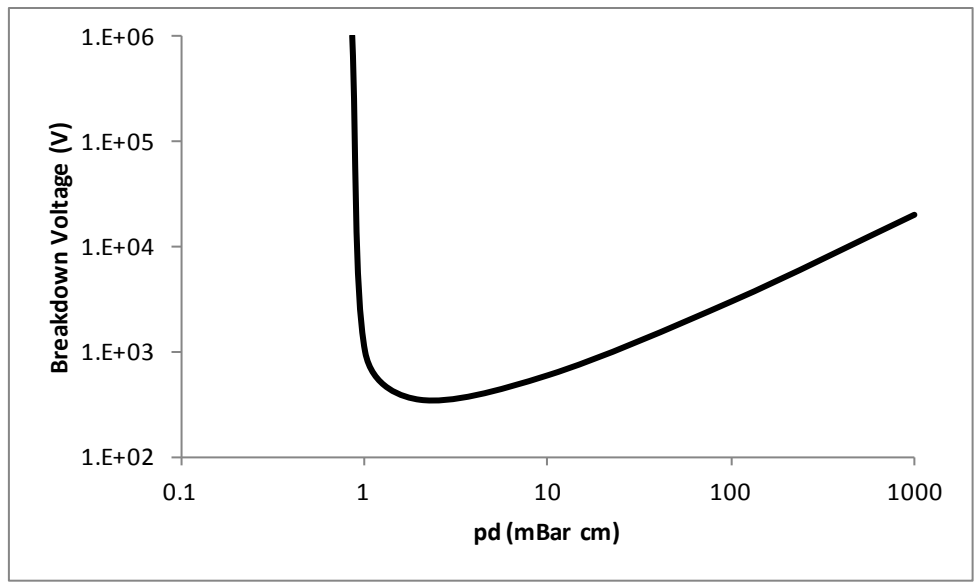

**Fig. 10:** The Paschen curve for hydrogen

At very high pressures the mean free path is very short. This means that the electrons never have enough time to be accelerated to the ionization energy before hitting an atom or molecule. This means that the breakdown voltage is high at very high pressures.

At very low pressures, the mean free path between collisions is longer than the distance between the electrodes. So, although the electrons can be accelerated to ionizing energies, they are unlikely to hit anything other than the anode. This means that the breakdown voltage is very high at very low pressures. This is why vacuum is such a good insulating medium.

Between these two extremes is a minimum whose position depends on the type of gas and the electrode material. This 'Paschen minimum' leads to a counterintuitive phenomenon: operating just below this minimum, electrodes further apart will have a lower breakdown voltage than those closer together. This is because, in a longer gap, there is more space for the electron avalanches to develop.

Ion sources often operate within an order of magnitude of the Paschen minimum because gas is added to a vacuum to create plasma for particle creation. Ion source engineers should always keep this in mind when trying to diagnose unwanted breakdowns. Figure 10 shows the Paschen curve for hydrogen, the Paschen curves for all other gases are very similar, with slightly different Paschen minima, ranging from 0.5 to 2 mbar cm and 100 to 800 V.

## 3.8 Vacuum breakdown

### 3.8.1 Introduction

Avalanche and streamer breakdown only occur at pressures and gap lengths to the right of the Paschen minimum shown in Fig. 10. To the left of the Paschen minimum is vacuum breakdown territory. This side of the Paschen minimum, the voltage required to break down a gap rapidly increases by several orders of magnitude.

Vacuum breakdown is a very different type of breakdown: a type in which the electrode surface plays a critical role. For breakdown to occur, a conductive plasma channel must bridge the gap between the electrodes. There are not enough atoms left between the electrodes to produce enough charge carriers, so they must instead come from the electrodes themselves.

### 3.8.2 Mechanisms

A surface can never be completely flat. Small or microscopic protuberances always exist at some scale. These cause localized field enhancements. When the voltage applied between the electrodes is high enough, small points on the cathode surface begin to emit electrons through quantum tunnelling processes. This highly localized current flow causes rapid localized overheating of the first cathode protrusion to start emitting. The protrusion explodes, creating a plume of vapour that is ionized and can cause breakdown.

The electrode material and surface finish are very important for vacuum breakdown. Highly polished surfaces are the best, but even they have a limit. The shiny smooth surface might still have microscopic oxide deposits in inclusions on the surface. The oxide deposits will generally be non-conductive, so they will have a permittivity different from vacuum. This step change in permittivity at the inclusion/oxide perimeter causes localized field enhancements. These localized field enhancements cause electrons to be emitted from the electrode. The resultant localized high current densities lead to a micro-explosion of material that is ionized, potentially causing breakdown.

### 3.8.3 High-voltage conditioning

High-voltage conditioning is essential to operating vacuum systems at very high field strengths. The conditioning process is very simple, but requires patience. The voltage is very slowly increased until the first sign of a discharge. The voltage is then held steady, or reduced slightly. When the discharge activity dies away, the voltage is very slowly increased again, until the discharge reappears. The voltage is then held steady, or reduced slightly. This process is repeated until the desired voltage is reached, or until a voltage is reached where the discharge activity does not die away.

Discharge activity might be detected through a sudden increase, or spike, in the supply current. If there is a window in the vessel, light might be visible, or if there is an antenna, electromagnetic noise could be picked up. Electrons emitted from the discharge are accelerated by the field and, if the voltage is high enough, could produce X-rays by bremsstrahlung. Depending on the configuration, there might not be any indication of a discharge until a complete flashover and resulting voltage collapse.

Whichever way the discharge is detected, the conditioning process works by slowly destroying each surface emitter in a controlled manner. If the voltage is increased too quickly, the discharge activity is too violent: instead of removing the protuberance, the localized electrode explosion creates more surface irregularities. These additional protuberances caused by the explosion might actually make the situation worse. The key to high-voltage conditioning is patience: the voltage must be increased very slowly, often being held at a voltage for some time before attempting to increase it further.

High-voltage conditioning is sometimes referred to as 'spot knocking' for obvious reasons.

Depending on the maximum field strength required, the conditioning process could take minutes or days. Below 1 MV m$^{-1}$ d.c., the conditioning process is relatively fast. Between 1 and 10 MV m$^{-1}$ d.c., the conditioning process takes longer. The best way to monitor how conditioning is going is to record the number of transient discharges (or sparks) per hour. At very high fields the spark rate might never get better than a few sparks per hour. The tolerable spark rate depends on the application. If no sparks can be tolerated, then the system must first be conditioned to a higher field, then when the voltage is reduced to the operating level the spark rate drops almost to zero. For very high field strengths above 10 MV m$^{-1}$ d.c., it is very difficult to condition the electrodes to give a spark rate of zero. The electrode material becomes very important at these sorts of field levels (see section 4.2.4).

## 3.9 Insulator breakdown

### 3.9.1 *Surface breakdown*

Something non-conducting has to hold the electrodes in position. This introduces an insulating surface between the electrodes. Insulator surfaces are the weakest point of any high-voltage system and care must be taken in their design (see section 4.3).

Discharges tend to propagate preferentially along surfaces because the surface is a good source of material to ionize and produce electrons. This allows surface avalanches to propagate along insulator surfaces in relatively low fields.

Triple junction effects (see section 4.3.2) where the insulator and the electrode meet can cause field enhancements, which can initiate discharges. Once started, the surface allows the discharge to propagate further than they would without the surface present.

The insulating surface can hold charge from previous discharge activity. Dependent on polarity, this can cause field enhancements, which can initiate surface discharges at lower voltages.

### 3.9.2 *Tracking*

Solid insulators are not self-healing like air and liquids. If there is a high-voltage surface breakdown, it can (but not always) cause permanent damage to the surface of the insulator. The high current that flows causes localized heating, leading to melting, carbonization or chemical changes of the insulator material and surrounding gases. This process is known as tracking: a conducting channel grows from one of the electrodes. It can be quite slow, building up a conducting track over a period of years, or it can be very quick, tracking over in one discharge event. The rate of growth depends on the voltages applied, the insulator material and its cleanliness. Once an insulator surface has completely tracked over, the electrodes are shorted together with a relatively low resistance, making application of a high voltage impossible. Some materials are track-resistant (in that they do not carbonize or form conductive compounds); these tend to be materials that do not contain hydrocarbons.

*3.9.3   Bulk breakdown*

Unlike its surface, the body of an insulator has a very high dielectric strength. The dielectric strength of a material is the field at which it fails or breaks down through its bulk. Gases and liquids also have dielectric strengths. In a solid, breakdown is always catastrophic; the material becomes punctured or ruptured with a track. Depending on the material, the track might be carbonized – it depends on what chemistry occurs when the material is turned into plasma. The shape of the track is defined by the shape of the electric field and the nature of the voltage applied. A uniform field will cause an almost straight track. A divergent field will cause a beautiful tree-like track as shown in Fig. 5.

The bulk of an insulator can also fail if it contains any voids. The effect of a void is twofold: firstly, the reduced permittivity in the void increases the field in the void; and secondly, the void is likely to contain a gas with a much lower dielectric strength than the insulator. Partial discharges in voids will eventually cause failure by slowly eroding their way through the insulation.

## 3.10   Partial discharges

Partial discharges are small high-voltage breakdowns that do not completely bridge the gap between the electrodes. They are caused by field enhancements created by changes in permittivity. This can either be voids in solid insulation (as discussed in the previous section) or mixtures of different insulator types.

The electric field in a void is higher because the permittivity of the void is lower than that of the surrounding bulk insulator. This causes the equipotential field lines to be concentrated in the void. Another way of looking at this is shown in Fig. 11(a). Consider that the void has a capacitance $C_v$, and the capacitance of the void is $C_x$ to one electrode and $C_y$ to the other. This makes a capacitive circuit between the electrodes, which can be thought of as a capacitive voltage divider. The void has a lower permittivity, and so a smaller capacitance. In a capacitive voltage divider the largest voltage drop is across the smallest capacitance, i.e., the void.

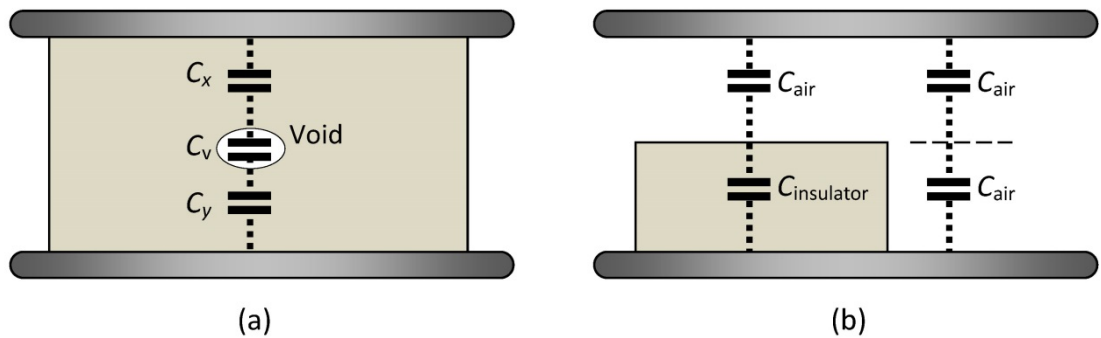

**Fig. 11:** Capacitive voltage divider explanation for how voids can cause field enhancement

The capacitive voltage divider analogy can also be used to explain why it is never a good idea to partially fill the gap between two electrodes with insulation. To increase the breakdown voltage between two electrodes, it would seem like a good idea to put some insulation between the electrodes. However, partially filling the gap, as shown in Fig. 11(b), actually increases the field in the region not filled with insulator. The capacitive voltage divider set up by $C_{air}$ and $C_{insulator}$ will make the field in the air greater than it would be without the insulator present! The top electrode will break down to the insulator surface. This could be a partial discharge, or it could go on to puncture the solid insulation.

To increase the breakdown voltage between electrodes, the only option is to completely fill the gap between them with a material of higher dielectric strength, making sure that no voids are present.

This is why vacuum injection moulding is often used to make solid high-voltage insulators because this avoids the presence of bubbles or voids.

Composite insulators can be built, but special care must be taken to understand the voltage distribution in them caused by different material permittivities.

**3.11 Corona discharges**

*3.11.1 Introduction*

A corona discharge is another type of partial discharge. They occur when there is a rapid change in field strength near a sharp point. Although this type of discharge usually falls into the 'unwanted' discharge territory, they can be tolerated. However, like partial discharges in voids, corona discharges generate electrical noise, which may be a problem in some situations. If they occur in air, they generate ozone, which can be harmful to health if there is inadequate ventilation. Corona discharges are used for some engineering applications, such as printing and filtration.

*3.11.2 Mechanism*

Figure 12 shows an electrode arrangement that will create a corona discharge before the electrodes flash over. Geometric field enhancement occurs near the tip of the triangle. The point causes the equipotential lines to bunch up, making the very localized high-field region shown in Fig. 12(b).

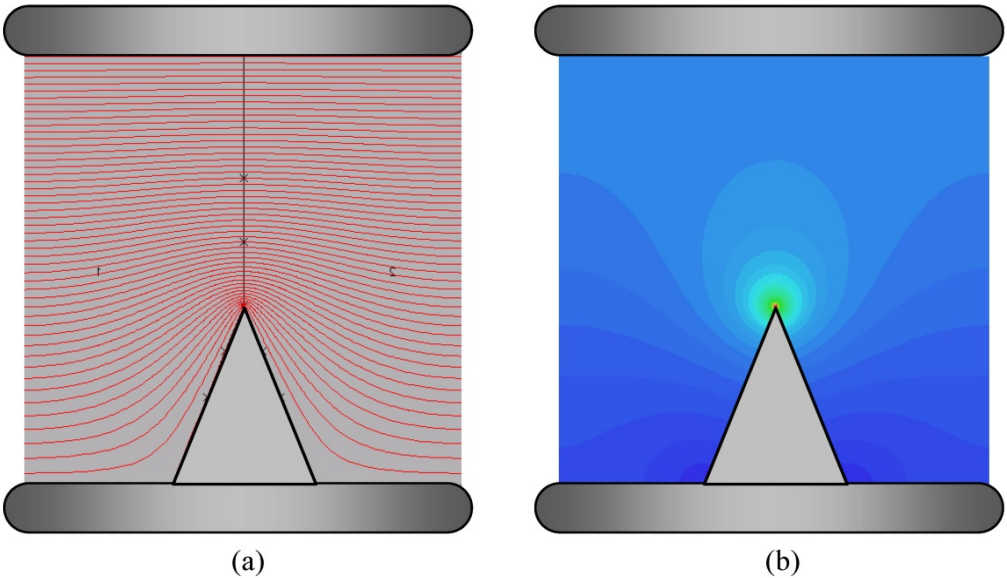

(a)                          (b)

**Fig. 12:** (a) Equipotential lines and (b) electric field strength around a point

When the applied voltage is high enough, discharges start in the high-field region at the tip and propagate into a decreasing field. The field strength drops off so quickly that the discharge can only propagate a short distance before the field is not strong enough to accelerate electrons to the ionization energy of the gas. The field in the rest of the gap is not strong enough for the discharges to develop either. The electron avalanches stop a certain distance from the tip.

*3.11.3 Importance of polarity*

In highly divergent fields like that shown in Fig. 12, the polarity of the applied field plays a critical role in the properties of the discharge. This is obvious when looking at the direction of the avalanches shown in Fig. 13.

For a positive point, shown in Fig. 13(a), the electron avalanches develop into an increasing field. Any avalanche that is initiated is guaranteed to develop until it reaches the positive point.

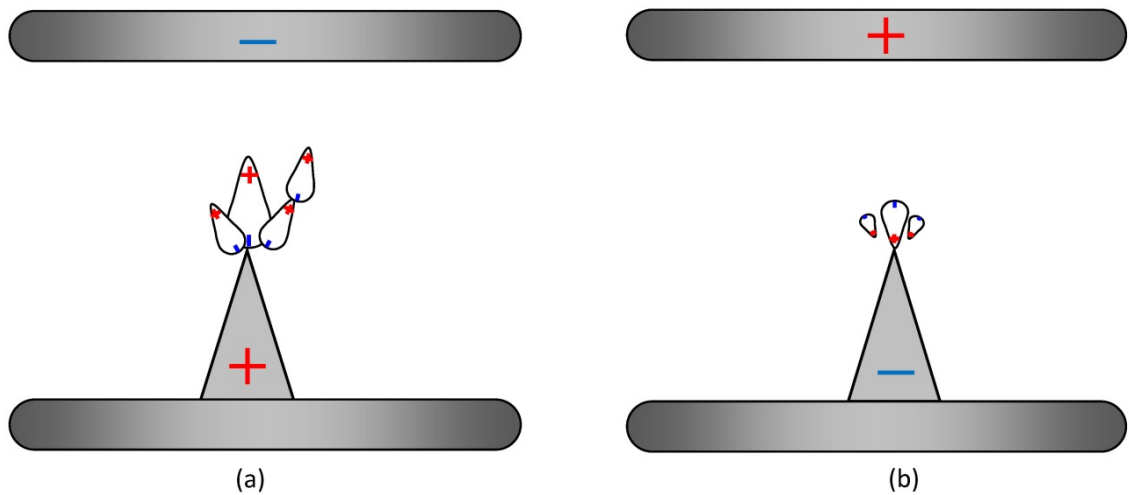

**Fig. 13**: Electron avalanches for (a) positive and (b) negative point–plane discharges for the same applied voltage.

For a negative point, shown in Fig. 13(b), the electron avalanches develop into a decreasing field. Avalanches will stop propagating when the field strength drops below the value required to accelerate the electrons to the ionization energy of the gas.

The asymmetry of the field means that discharges produced from a positive point propagate much further into the gap than those from a negative point at the same voltage. The field enhancement caused by the charge separation in the initial avalanches causes secondary avalanches. Positive point discharges shown in Fig. 13(a) are more likely to cause secondary avalanches because the size of the initial avalanches is much larger.

*3.11.4 Corona inception and extinction voltage*

The voltage at which the corona activity starts is referred to as the inception voltage. The field asymmetry in a point–plane gap also means that the voltage at which the corona discharge starts is different for each polarity. Once started, the corona discharge will continue until the applied voltage is reduced. Avalanches form on a nanosecond timescale, and new avalanches are continuously developing near the point.

If the applied voltage is reduced, the corona discharge eventually stops. The voltage at which this happens is referred to as the corona extinction voltage. It is always lower than the inception voltage because, once started, there are always lots of free electrons present to produce more avalanches. Extra free electrons are also created for a negative point corona from electron emission caused by ion bombardment.

**3.12 Importance of the power supply**

The power source attached to the electrodes will have a large effect on the type of discharge produced. If it is an a.c. supply, the voltage applied to the electrodes will change polarity each cycle, which will cause the discharge to start and stop twice each cycle. If it is a pulsed supply, there may be time for only a few avalanches before the voltage is switched off. The amount of energy stored in the supply will limit how much current flows before the voltage collapses and the discharge stops. If the power source has a large capacitance (or large inductance), then there could be enough stored energy to cause catastrophic damage to the electrodes and insulators. If the power supply is capable of sustaining an 'unwanted' discharge, it is vital that discharge detection systems can effectively inhibit the power supply output before serious damage occurs by arcing (section 3.14).

All the discharges discussed so far fall into the category of 'unwanted' avalanche discharges. Looking back to the landscape of different discharges shown in Fig. 6, the regimes for 'wanted' discharges occur after the point of avalanche breakdown.

The current–voltage characteristics of discharges are not ohmic. In fact, the arc discharge regime shown in Fig. 5 (discussed in section 3.14) actually has a negative resistance: the volts drop as the current increases.

For 'wanted' discharges the supply must be able to deliver power reliably to the discharge and provide stable control of the current flowing. The current and voltage of a discharge will be where the power supply load curve intersects the characteristic shown in Fig. 6. The gradient of the discharge characteristic at the intersection point determines whether the discharge is stable or not. A resistor is often put in series with the discharge to allow stable control of the current.

### 3.13 Glow discharge

#### 3.13.1 *Introduction*

Glow discharges are used in many applications: fluorescent tubes, neon tubes, plasma televisions, for surface treatment and, of course, in ion sources. The glow discharge is so called because it emits a significant amount of light. Most of the photons that make up this light are produced when atoms that have had their orbital electrons excited by electron bombardment relax back to their ground states. Photons are produced in any event that needs to release energy, for example when ions recombine with the free electrons and when vibrationally excited molecules relax.

#### 3.13.2 *Striking a glow*

The process of starting a glow discharge is referred to as striking. The gas between the electrodes must be ionized. This is usually achieved by simply applying a voltage high enough to cause breakdown. The runaway avalanche breakdown process causes the voltage needed to sustain a discharge to collapse to some tens of volts, as shown in Fig. 6. This collapse in voltage signifies that the discharge has entered the glow discharge regime. The gas between the electrodes has become highly conductive plasma.

When struck, a glow discharge is self-sustaining because positive ions are accelerated to the cathode, and when they impact on it, they produce more electrons. This is one of the secondary emission processes, $\gamma_{\text{ion}}$, discussed in section 3.5.3. The self-sustaining nature of glow discharges is why there is a hysteresis in the current versus voltage curve at the glow-to-dark discharge transition shown in Fig. 6.

#### 3.13.3 *Debye length*

Named after the Dutch scientist Peter Debye, the Debye length $\lambda_{\text{D}}$ is the distance over which the free electrons redistribute themselves to screen out electric fields in plasma. This screening process occurs because the light mobile electrons are repelled from each other while being pulled by neighbouring heavy low-mobility positive ions. Thus the electrons will always distribute themselves between the ions. The electron electric fields counteract the fields of the ions, creating a screening effect. The

Debye length not only limits the influential range that particles' electric fields have on each other but also limits how far electric fields produced by voltages applied to electrodes can penetrate into the plasma. The Debye length effect is what makes the plasma quasi-neutral over long distances.

The higher the electron density, the more effective the screening, and thus the shorter this screening (Debye) length will be. The Debye length is given by:

$$\lambda_\mathrm{D} = \sqrt{\frac{\epsilon_0 k T_\mathrm{e}}{n_\mathrm{e} q_\mathrm{e}^2}} \qquad (34)$$

where

$\lambda_\mathrm{D}$ is the Debye length (of order 0.1–1 mm for ion source plasmas),

$\epsilon_0$ is the permittivity of free space,

$k$ is the Boltzmann constant,

$q_\mathrm{e}$ is the charge of an electron,

$T_\mathrm{e}$ is the temperatures of the electrons,

$n_\mathrm{e}$ is the density of electrons.

### 3.13.4  Cathode plasma sheath

The screening effect of the plasma creates a phenomenon called the cathode plasma sheath around the cathode electrode. The plasma sheath is also called the Debye sheath. The sheath has a greater density of positive ions, and hence an overall excess positive charge. It balances an opposite negative charge on the cathode with which it is in contact. The plasma sheath is several Debye lengths thick.

The majority of the voltage between the anode and the cathode is dropped across this thin cathode plasma sheath. The bulk of plasma is almost at the anode potential.

### 3.13.5  Towards arc transition

The current in a glow discharge can be increased over several orders of magnitude with very little increase in discharge voltage. The plasma distributes itself around the cathode surface as the current increases. Eventually the current reaches a point where the cathode surface is completely covered with plasma and the only way to increase the current further is to increase the current density at the cathode. This causes a larger voltage drop near the cathode and the plasma voltage rises as shown in Fig. 6.

## 3.14  Arc discharge

### 3.14.1  Introduction

Arcs are used in welding, cutting, machining, furnaces and ion sources. Unwanted arc discharges can be very destructive, causing significant damage to electrodes and insulators by sputtering and melting from the large currents that flow. They can also destroy power supplies because of their negative resistance characteristic shown in Fig. 6. More current flows as the power supply volts drop. Without adequate protection, the components in the power supply output stage will fail due to overcurrent. The current in an arc is only limited by what the power supply can deliver.

### 3.14.2  Glow to arc transition

If the power supply is capable of driving an arc, then arc discharges can occur directly after high-voltage breakdown. The current increases so rapidly that there is no time for a glow discharge to form.

The current density at the cathode causes heating and eventually the cathode surface reaches a temperature where it starts to emit electrons thermionically. This is technically when the discharge moves into the arc regime with a negative current–voltage characteristic as shown in Fig. 6.

The transition between glow and arc can be unstable because of the rapid increase in emitted electrons with temperature and the dynamic surface processes.

Magnetron and Penning ion sources that operate in the arc discharge regime enhance the electron emission from the cathode with the addition of work-function-lowering caesium. Vacuum arc ion sources rely on the heating and vaporizing of the cathode material to produce ions.

### 3.14.3 Thermal arcs

As the current in an arc increases, the plasma becomes almost completely ionized (there are few neutral particles left). Eventually the current density in the plasma reaches a point where the ions have the same average velocity as the electrons: they have reached thermal equilibrium. The discharge enters the thermal arc regime where the discharge voltage rises as the current increases, as shown in Fig. 6.

## 3.15 Factors affecting properties of discharges

### 3.15.1 Statistical variability

Even with identical conditions, the same electrode gap will break down at different voltages each time the voltage is applied. This is because of the statistical nature of high-voltage breakdown: no two sparks are ever the same. The initial location of the free electrons will be different every time.

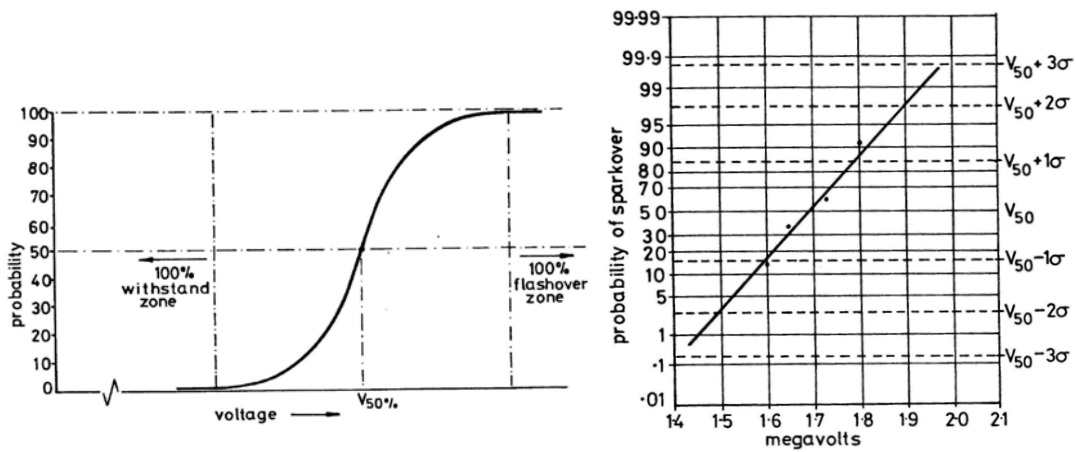

**Fig. 14:** The statistical variability of high-voltage breakdown

To help to quantify this, the concept of $V_{50\%}$ is introduced. This is the voltage at which the system will break down 50% of the time the voltage is applied. Figure 14 illustrates this point. It also demonstrates that operating at a voltage just below $V_{50\%}$ does not guarantee that breakdowns will not occur: they are just less likely. For applications where high-voltage breakdowns cannot be tolerated, significant safety margins must be introduced.

The value of $V_{50\%}$ can only be obtained by repeated experiment, but this may not be possible in certain circumstances. For example, an insulator might be permanently damaged after a single breakdown. High-voltage conditioning of vacuum systems is essentially a process of carefully finding the $V_{50\%}$ then slowly increasing the voltage until $V_{50\%}$ can no longer be increased.

### 3.15.2 Environmental conditions

Higher temperatures and lower pressures lead to lower flashover voltages. A correction factor for $V_{50\%}$ in air can be found from

$$V_{50\%} \text{ correction factor} = \frac{0.386 \times P}{273 + t} \tag{35}$$

where $P$ is in mmHg and $t$ is in degrees Celsius.

The humidity of air can also affect the breakdown voltage. Higher humidity leads to lower breakdown and corona inception voltages.

### 3.15.3 Polarity and geometry

The influence that voltage polarity has on discharges has already been discussed in section 3.11. If there is perfect mirror symmetry between the anode and cathode, the polarity of the applied voltage does not matter. As soon as the mirror symmetry is broken, the positive and negative breakdown voltages of the geometry become different. The difference in breakdown voltage is largest when the difference between the anode and cathode geometry is largest. The most extreme case is the point–plane gap shown in Fig. 12. In this case a negative point breaks down at a much higher voltage than a positive point.

### 3.15.4 Type of applied voltage

The rise-time and duration of the applied voltage will affect at what voltage the gap breaks down. Shorter applied pulses, or higher-frequency a.c., have a higher breakdown voltage than d.c. or slow pulses. If the applied voltage pulse is short enough, the discharge does not have time to build up before the voltage drops.

### 3.15.5 Magnetic fields

Magnetic fields significantly affect electrical discharges. Charged particles will rotate around magnetic field lines: this means that they tend to travel along magnetic field lines, by spiralling along them. This can be exploited to confine plasma in an ion source.

Electrons are much lighter than ions, so magnetic fields have a much greater effect on electrons than they do on ions. Stray magnetic fields can cause unwanted breakdowns by channelling electrons along the field lines.

### 3.15.6 Charges

Discharges by their very nature contain large numbers of charge carriers. The space charge of the free charge carriers created by a discharge can have a significant effect on the overall electric field. Depending on the polarity and timeframes of the applied voltages, this can either enhance or reduce the applied field. The same effect can happen when surface charge is deposited on insulator surfaces. Residual charge left by a previous discharge can alter the behaviour of the following discharge. The overall effect that charges have on discharges is complicated and situation-dependent. Residual charges are a cause of some of the statistical variability of discharges.

### 3.15.7 Contamination and lost beams

The cleanliness of the electrode and insulator surfaces can reduce the breakdown voltage of a system. Pollution, dirt, grease and water can all cause tracking (section 3.9.2) to occur. Figure 15 shows a 35 kV insulator that has tracked by water flowing down its surface. Stray particle beams impacting on insulators or electrodes can trigger breakdowns. In vacuum, X-rays from bremsstrahlung created when electrons hit electrode surfaces can cause insulators to break down.

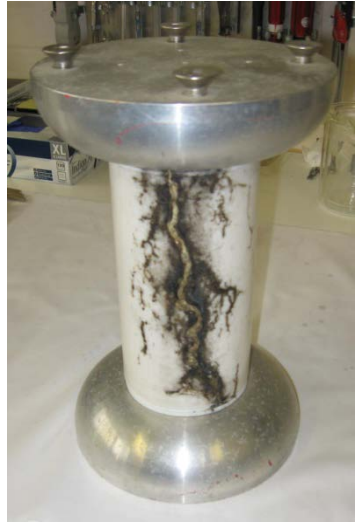

**Fig. 15:** Tracking caused by water leak across a 35 kV insulator

## 4 High-voltage design and technology

### 4.1 Introduction

This section covers some of the design principles required to successfully design and build high-voltage systems. It also discusses the commercial high-voltage technology required to bring complete ion source systems together.

### 4.2 Electrodes

#### *4.2.1 Electrode design*

Electrodes are used to shape the electric field or as a source/sink for current. Field shaping electrodes are used in accelerating, deflecting and focusing beams. Current source/sink electrodes are used in plasma generation and beam dumps. Some electrodes such as extraction electrodes can be both field shaping and current sinking.

Electrode design obviously depends on application. For field shaping electrodes, the shape of the electrode is critical. For current source/sink electrodes, the sputtering resistance and current-carrying capability is critical.

If the electrode's aim is to shape fields while avoiding unwanted discharges, the key to high-voltage electrode design is smoothness, both microscopically and macroscopically. Electrodes should be physically smooth and have gently curving features with large radii.

Using modern finite element techniques, the electric fields produced by electrodes can be accurately calculated (see section 2.6). This allows designs to be tested and optimized before manufacture. Extraction and beam acceleration electrodes can be modelled and simulated beams tracked through them.

*4.2.2 Corona shields*

Corona shields are a type of field shaping electrode used to shield sharp points that would otherwise suffer from corona. The sharp points are usually unavoidable components like bolts or other fixtures and fittings. A corona shield increases the radius of these points by literally covering them up with rounded electrodes that are electrically connected so they are effectively at the same potential. Other names for corona shields are guard rings and stress rings. Figure 16 shows two examples of how to shield a point with two different types of corona shield.

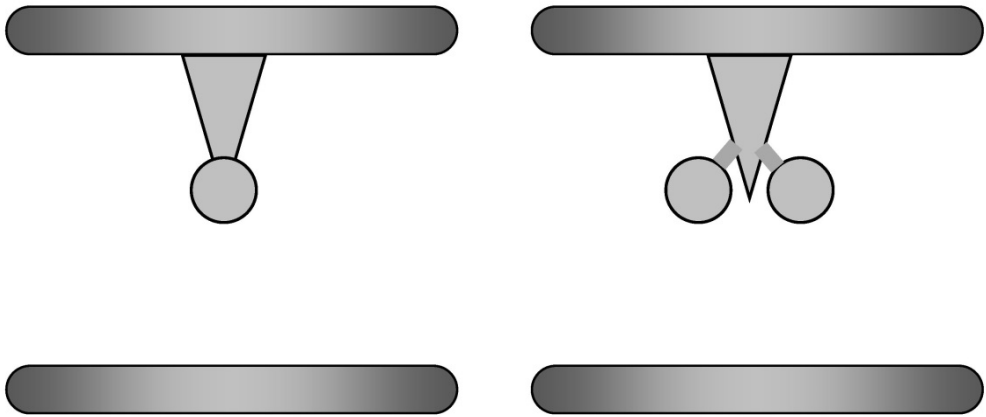

**Fig. 16:** Two examples of shielding a point with corona shields

*4.2.3 Sputtering*

Sputtering is where ions bombard electrode surfaces, knocking atoms off the surface. This process causes erosion of the electrode surface. Electrodes in plasmas suffer the most from sputtering. Sputtering eventually leads to failure of the plasma electrode system. The process also slowly deforms extraction electrodes.

*4.2.4 Electrode materials*

For cathodes in high-sputter environments, molybdenum or tungsten are often used. For field shaping electrodes, any conductive material can be used, so material choice is application-dependent.

For higher fields, electrodes must be smooth and highly polished to minimize surface irregularities. Polished 316L stainless steel is commonly used for most applications.

For operation at very high field strengths in vacuum >10 MV m$^{-1}$ d.c., titanium or its alloys give lower spark rates, especially when used as the cathode electrode material. Titanium–aluminium–molybdenum alloy and pure titanium produce approximately equal spark rates. Titanium–vanadium or titanium–manganese alloys have the lowest spark rates of all [5].

**4.3 Insulators**

*4.3.1 System design*

Insulator design is based on the fact that an insulator's surface is the weakest part of a high-voltage system. Good design must: keep the electric field strength along its surface as low as possible; avoid field enhancement caused by triple junctions; and keep the surface clean.

Breakdowns on insulator surfaces cause permanent damage, so good HV system design must ensure that breakdowns are unlikely to occur across an insulator surface. The highest field strengths should be across electrode-to-electrode gaps. The field strengths along insulator surfaces should be kept as low as possible by making the insulators as long as is practically possible. Triple junctions that could initiate a discharge should be screened.

### *4.3.2   Triple junction effect*

The triple junction effect is a localized geometric field enhancement caused when insulators of two different permittivities meet at an electrode. Triple junctions always exist at some scale because it is impossible to have a completely flat surface (see Fig. 17).

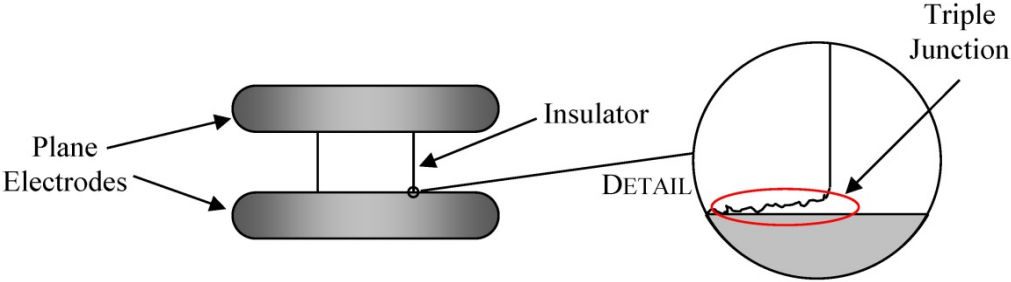

**Fig. 17:** Triple junctions occur where insulators and electrodes meet

The triple junctions cause field enhancement because free space (air or vacuum) has a lower permittivity than the insulator. The insulator pushes the equipotential field lines into the free space, increasing the field strength in that region. An alternative capacitive voltage divider explanation is given in section 3.10.

Triple junctions are a problem because the field enhanced region is in the perfect position to initiate discharges, which can then propagate across the entire insulator surface. The field enhancement effect actually gets worse the smaller the triple junction becomes. This is illustrated in Fig. 18, which shows how the field in the free space region increases as the triple junction gap is reduced from 4 mm to 0.5 mm in a 500 kV m$^{-1}$ ambient field strength with an insulator relative permittivity of 2.2. At 0.5 mm the field in the free space region has almost doubled to 1000 kV m$^{-1}$.

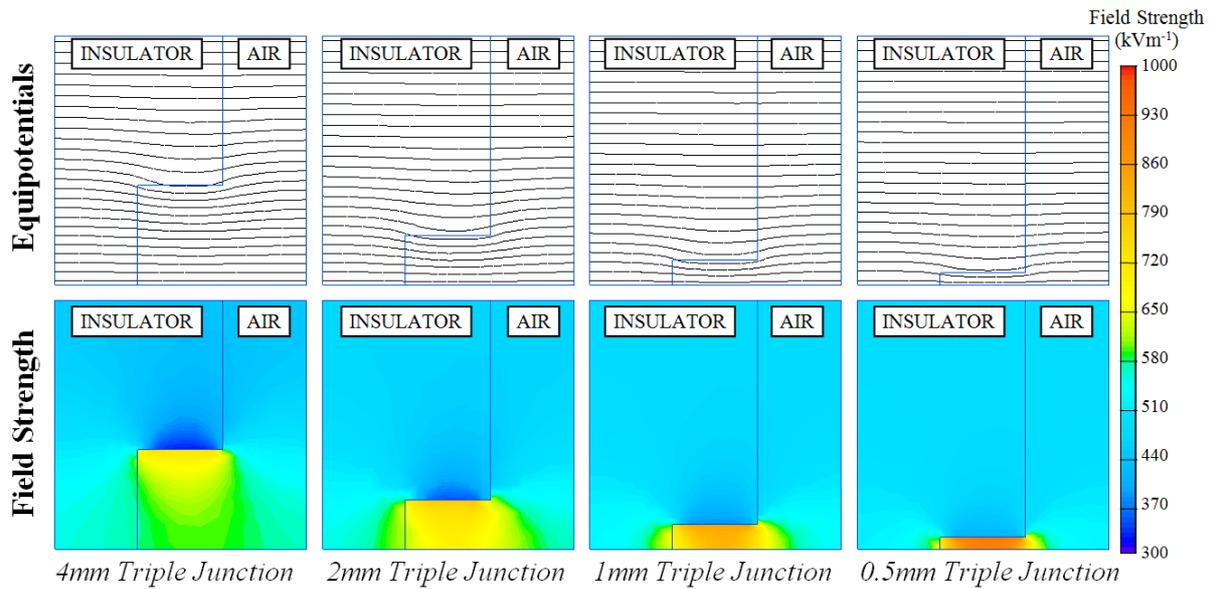

**Fig. 18:** Triple junction field enhancement increases as the gap become smaller

The permittivity of the insulator amplifies the triple junction effect because the increased relative permittivity pushes more field lines into the free space. Figure 19 shows how the field in a 1 mm triple gap increases as the relative permittivity of the insulator increases in a 500 kV m$^{-1}$ ambient field.

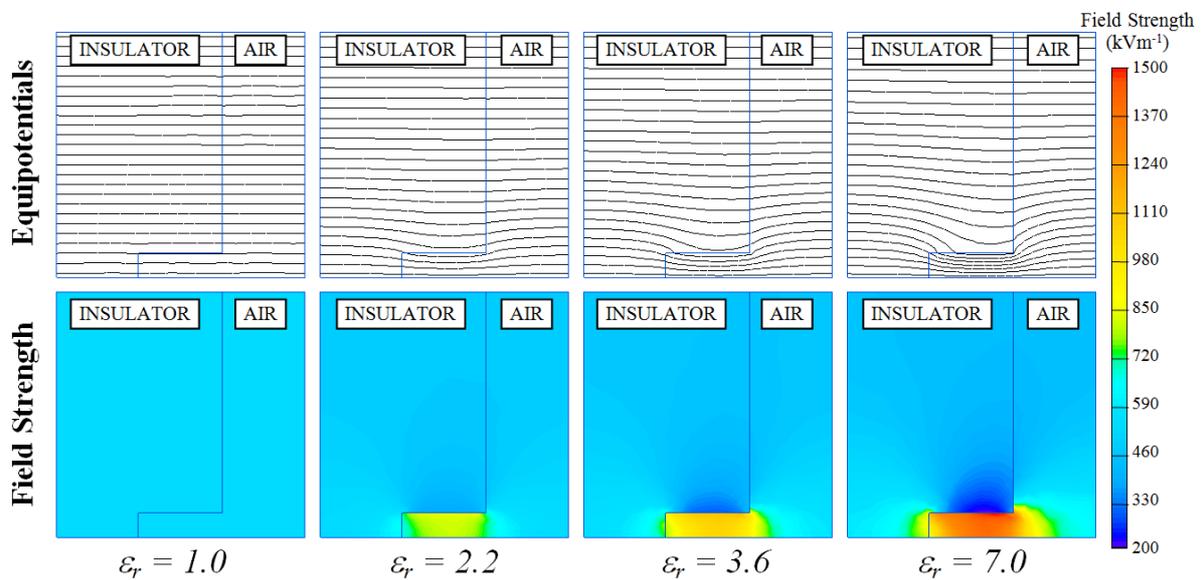

**Fig. 19:** Triple junction field enhancement increases as the permittivity of the insulator increases

### 4.3.3 *Triple junction shielding*

By shaping the electrodes, it is possible to shield any triple junction. Figure 20 shows some of the different possible techniques, which all work by reducing the field in the triple junction region.

Fig. 20(a) shows the field in an insulator with a relative permeability of 2.2 and a 1 mm triple junction in an ambient field of 500 kV m$^{-1}$.

If the insulator cannot be modified, then the best option is to recess the electrode as shown in Fig. 20(b). If neither insulator nor the electrode can be modified, then the best option is to add a guard ring onto the electrode, as shown in Fig. 20(c). Both the guard rings and the recesses need to be a short distance away from the insulator: close enough to shield the triple junction, yet far enough away so the field on the insulator surface is not increased. The precise shape of the guard ring cross-section can be optimised to minimise the field on the insulator surface.

If the insulator can be modified, the best solution is to add an internal guard ring, as shown in Fig. 20(d). This reduces the field in free space and increases it in the insulator material. This is the best engineering solution because the bulk of the insulator has a very high dielectric strength. Also the conditions inside the insulator are much more stable and controllable than outside. Surface contamination and lost beams can cause the surface of an insulator to fail (section 3.15.7). It is also possible to reduce the surface field even further by also adding an external guard ring or recess, as shown in Fig. 20(e).

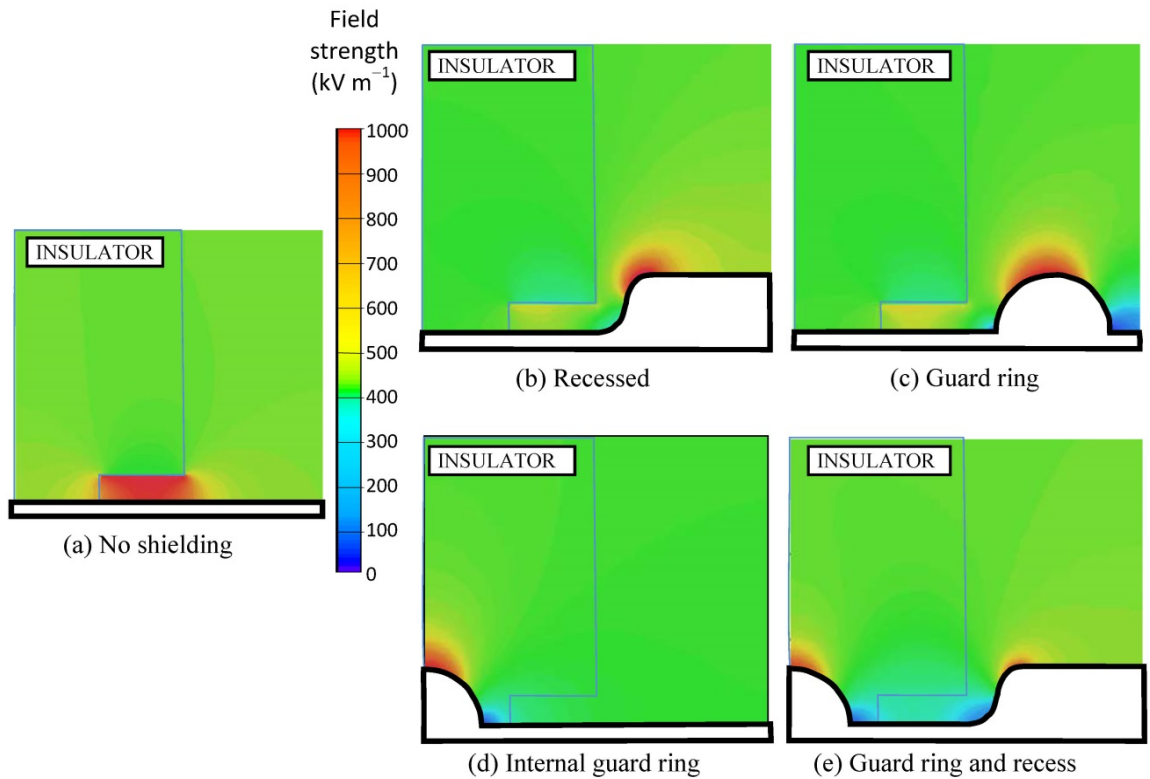

**Fig. 20:** Different insulator triple junction shielding techniques

### *4.3.4    Insulator material*

There are many types of insulation material. Which insulator to use obviously depends on the application. For example, $Al_2O_3$ is commonly used in vacuum. AlN is used when high thermal conductivity is required. Macor is used when small complex shapes need to be machined. Porcelain is used in compression. Epoxy resin is used to impregnate and pot. Mica is used for thin high-voltage withstand. Glass is used when visible transparency is required.

Table 1 shows the relative permittivity and dielectric strength of some commonly used insulation materials.

**Table 1:** The relative permittivity and dielectric strength of some commonly used insulation materials.

| Material | Relative permittivity, $\varepsilon_r$ | Dielectric strength (kV mm$^{-1}$) |
|---|---|---|
| Air | 1 | 3 |
| PTFE | 2.2 | 19.7 |
| Al$_2$O$_3$ | 8–10 | 13.4 |
| Mica | 6 | 118 |
| Epoxy resin | 3.6 | 20 |
| SF$_6$ | 1.002 | 7.5 |
| Oil | 2.5–4 | 10–15 |

### 4.3.5  Insulator surface profile

If the electric fields along a surface are low ($\ll 5$ kV cm$^{-1}$), then simple, straight, smooth insulators can be used in many applications. However, for high-field applications, the shape of an insulator's surface profile is a very important part of its high-voltage design.

The ridged or contoured profile that is commonly seen on high-voltage insulators is there to increase the surface path length along which the discharge must propagate to reach the other electrode. The profile should also make the electric field component parallel to the surface drop to zero along parts of the surface, which will inhibit surface propagation.

Another name for the insulator ridges is 'sheds'. This name is more common in insulators designed to work outdoors and in wet conditions. Water flowing down insulator surfaces can cause tracking (see section 3.9.2) in fields much lower than 5 kV cm$^{-1}$. The field strength that caused the track shown in Fig. 15 was only 1.2 kV cm$^{-1}$. Sheds work by protecting some of the insulator's surface from the water, keeping it dry like a garden shed keeps its contents dry. In extreme environments, the underside of the shed has additional undulations. The sheds are designed so that water cannot collect on their surfaces. The water flowing off the surfaces has the additional benefit of cleaning off any pollution.

### 4.3.6  Insulator protection

In ion sources that use caesium or sources that sputter large amounts of conductive metals, it is sometimes necessary to implement insulator shields around the insulator. These are metal shields surrounding, but not touching, the insulator. They are mounted on one electrode, and stop short of the other electrode. They leave a small section of insulator visible, with the majority of it covered. If the environment is particularly harsh, it is sometimes necessary to add a second shield of larger diameter connected to the other electrode, which covers both the insulator and first insulator shield.

When an insulator flashes over, it can be irreparably damaged. To prevent this, extra electrodes can be added that act as a protective spark gap. Called 'arcing horns', they limit the voltage that can be applied to the insulator, ensuring that it does not become damaged.

### 4.3.7  Insulation coordination

When looking at the overall system design, it is important to make sure that each part of the high-voltage system is capable of working at the voltages required. Starting at the high-voltage power supply, the cable connecting it to the vacuum vessel should be able to withstand the maximum output voltage. The vacuum feedthrough should be able to withstand the maximum output voltage. Any insulators inside the vessel should be able to withstand the maximum output voltage. It should only be the electrodes themselves that might have a lower flashover voltage.

### 4.4 Gaseous and liquid insulation

#### *4.4.1 Air*

Air is obviously the most common form of gaseous insulation. As a rule of thumb, air at normal ambient conditions requires about 30 kV cm$^{-1}$ to break down, for uniform fields. This number slowly drops for longer gaps and is significantly lower for asymmetrical fields. In fact, a positive point–plane 1 m gap only requires 500 kV to break down.

The operating voltage for a high-voltage system should always be significantly lower (see Fig. 13) than its breakdown voltage if the system is to operate reliably.

#### *4.4.2 Sulphur hexafluoride (SF$_6$)*

SF$_6$ is the most commonly used insulating gas. It has a dielectric strength twice that of air at one atmosphere pressure because it is a very electronegative gas. This makes it very good at absorbing the free electrons produced in avalanches. It thus inhibits electrical breakdown.

It is non-toxic and non-flammable, making it very safe to work with. It allows lightweight, compact, high-voltage systems to be developed without the need for vacuum pumps. Often used in vessels filled to only 0.5 bar over atmospheric pressure, it can easily deliver dielectric strengths three times that of air. For specialist applications, even higher pressures are used.

The major drawback of SF$_6$ is that it is the most potent greenhouse gas: 1 kg of SF$_6$ is equivalent to 23 900 kg of CO$_2$. It requires gas handling systems and must not be released into the atmosphere. If significant amount of sparking occurs, by-products can be produced that are hazardous to health and can be corrosive. The most toxic gaseous arcing by-product is S$_2$F$_{10}$, which has a permitted exposure limit of only 0.01 ppm.

#### *4.4.3 Oil*

Oil is the most common liquid insulator. It has a similar dielectric strength to that of SF$_6$ at 0.5 bar overpressure (i.e., three times that of air). Oil is commonly used in transformers and high-current applications because it can also be used as a coolant.

Oil is much less harmful to the environment, but it still must be disposed of properly, and a back-up containment vessel must be provided to prevent leakage. This is in the form of either double-skinned tanks or bunds.

### 4.5 Commercially available HV components

#### *4.5.1 Bushings*

Bushings are high-voltage feedthroughs designed to pass a high voltage through a barrier. For ion sources, this barrier is often a vacuum vessel, so the bushing must also withstand the pressure difference. The length of the bushing is much more compact on the vacuum side of the bushing than it is on the air side. Bushings are used for semi-permanent high-voltage connections.

#### *4.5.2 Connectors*

Connectors are used to connect high-voltage cables between the high-voltage systems and the ion source. Like bushings, they can also be mounted on a vacuum vessel. Connectors are used when the high-voltage connection needs to be regularly removed.

*4.5.3 Cables*

High-voltage cable is a specialized system in itself and it will be designed to operate at a certain voltage. High-voltage cable is usually coaxial, with a centre high-voltage conductor surrounded by an earthed outer conductor. The key to successful cable use is proper cable termination. High-voltage cables cannot just be stripped back like normal wires. The insulation must be carefully removed, and often other components such a stress cones or stress shields are required to screen the sharp edge created when the outer conductor was cut. Sometimes self-amalgamating high-permittivity tape is used to spread the electric field away from the sharp-edged outer conductor.

High-voltage connectors are designed for specific cables, so they include cable terminating features in their design.

*4.5.4 Insulators*

Ion sources contain lots of custom-built insulators designed to fit in specific places, but wherever possible commercially available insulators should be used. Commercially available insulators are relatively cheap and, if used according to their specifications, incredibly reliable. A well-designed insulation system is one you should never have to worry about.

*4.5.5 Voltage dividers*

It is necessary to measure and monitor the high voltages applied in a system. This is usually achieved with voltage dividers. These can be either capacitive or resistive. Some dividers that are designed to work with d.c. and give a flat frequency response are a combination of both.

**4.6 High-voltage platforms**

*4.6.1 Introduction*

Ion source technology nearly always requires the biasing of power supplies at high voltages. This is because electric fields are used to extract beams from plasma. To do this, the source plasma must be held at a high voltage, and the extracted beam accelerated to ground potential. To achieve this set-up, the power supply required to generate the plasma must also be biased at a high voltage. Usually the power supply needs some sort of control system, which also must be held at a high voltage, also called 'at volts'. In addition the ion source usually needs some sort of monitoring 'at volts'.

All this is achieved by electrically floating all the equipment on a high-voltage platform, powering them with isolated power supplies, and controlling them with isolated signals.

*4.6.2 System design*

There are two possible system design philosophies for high-voltage platforms.

*High-voltage 'room'*

This involves physically putting all the equipment on a metal platform supported by insulators. The connections to the ion source are made with wires capable of carrying the required currents and suitable vacuum feedthroughs. The ion source is mounted on a high-voltage insulator. The platform, connecting wires and ion source are all surrounded by a cage, room or partition walls to prevent people from touching equipment biased at volts.

This design philosophy takes up a lot of room. If electrical noise or X-rays are produced, then the entire 'room' must be adequately screened and shielded. The main advantage of a high-voltage 'room' is that it allows plenty of space for expansion. Temporary diagnostics and experimental systems can easily be accommodated and simply connected to the ion source. Access for maintenance is easy. For a source under development, this is the best design philosophy.

*High-voltage 'chassis'*

The equipment that needs to run at volts is mounted in a chassis. The chassis is mounted on insulators in an enclosure to prevent people from touching the high-voltage chassis. High-voltage cable is used to take the power and monitoring signals to the ion source. High-voltage feedthroughs then take the electrical connections into the vacuum vessel and onto the ion source.

This design philosophy is very inflexible. There may be no room in the HV chassis for any more equipment. Any new equipment will need new high-voltage cables and feedthroughs to the ion source. Equipment maintenance requires opening two enclosures, and access is often restricted. The main advantage of an HV chassis is compactness. All high voltages are safely enclosed, electrical screening is easy, and X-ray shielding can be implemented without difficulty if required. Personnel can have full access to the area while the source is running. For a mature, fully developed system, this is the best design philosophy.

Hybrid design philosophies are possible. For example, an enclosed HV chassis could be implemented with a caged cable tray connecting to the ion source. This removes the need for HV cabling to carry low-voltage power and monitoring signals. The main question the system designer needs to ask is this: How flexible does the set-up need to be?

### 4.6.3   *Isolated power supplies*

Isolated power supplies are used to get power to the equipment on the HV platform. The simplest solution is a 1 : 1 mains isolation transformer. An isolation transformer has a large amount of insulation between its secondary winding and the yoke. This allows the primary winding and the magnetic steel yoke to remain at ground potential, while the electrically isolated secondary can be floated to platform volts. Isolation transformers can be air- or oil-filled, and they can be single- or three-phase.

There is a limit to how much insulation can be wrapped around a transformer core. For most cases this is a few hundred kilovolts. Beyond this voltage, the only way to get power to the high-voltage platform is with a motor alternator set. A motor is mounted at ground potential and mechanically connected via an insulating rod of material to a generator mounted on the high-voltage platform.

The number of moving parts makes motor alternator sets much less reliable than isolation transformers. Whichever system is used, the isolated neutral must be connected to the high-voltage platform for safety. Otherwise, when the platform is earthed down, the platform power supply would not be referenced to anything.

### 4.6.4   *Isolated control signals*

There must be a way to control and monitor the equipment on the high-voltage platform. An old-fashioned technique is to use long insulating rods. This has now been almost entirely replaced by signals sent over fibre optics. There are many commercially available control systems that can communicate over fibre.

### 4.6.5   *Other isolated systems*

It is sometimes necessary to connect a water chiller at ground potential to the ion source on the high-voltage platform. To do this, long insulating hoses are used. The water circuit must be kept deionized with an ion exchange column. For safety it is a good idea also to have a conductivity meter, which provides an interlock signal to the high-voltage power supply.

Gases, including hydrogen can be delivered to the high-voltage platform with simple insulating tubes.

Radio-frequency (RF) power can be fed to the platform through specially designed waveguides with insulating breaks.

**4.7  High-voltage power supplies**

*4.7.1   Power supply technologies*

There are many different ways to generate high voltages:

- Semiconductors – thyristor, insulated gate bipolar transistor (IGBT), gate turn-off (GTO) thyristor
- Tube – tetrode
- Pulse-forming network
- Cascade rectifier (Greinacher, Cockcroft–Walton multiplier)
- Van de Graaff generator, pelletron
- Linear (usually front end only)
- Switched mode transformer – HV diode and capacitor

*4.7.2   High-voltage power supply manufacturers*

There are many manufacturers of high-voltage power supplies. Standard designs are available or custom-built power supplies can be procured. Tight specification is essential. Ion sources are an incredibly difficult load for a power supply to deliver power to. The power supply must be able to cope with breakdowns – this must be specified.

If the resource, ability and skills are available, it is possible build high-voltage power supplies 'in house'.

*4.7.3   Insulation test equipment*

High-voltage systems can be tested with special high-voltage power supplies that only allow very small currents to flow in the event of a breakdown. This allows systems to be tested with a reduced risk of permanently damaging any solid insulation.

**5   Safety**

**5.1   Introduction**

There are many hazards that must be considered when working with ion sources: high temperatures, strong magnetic fields, dangerous materials (e.g. caesium) and heavy equipment. Before any work starts, thorough risk assessments should be made. This section covers the safety hazards involved when working with high voltages.

## 5.2 Electric shocks

The main hazard from high voltage is electric shock. Small electric shocks can startle people, potentially causing them to fall or injure themselves by hitting or dropping something. We can sense small electric currents as low as 5 mA (either a.c. or d.c.) as a mild tingling sensation. Higher currents can cause the muscles to tense: 60 mA (either a.c. or d.c.) flowing across the chest cavity is enough to fibrillate the heart, potentially causing heart attack and death. Therefore, the most dangerous situation is holding an earthed handrail and touching a live conductor. A saying goes: 'An old electrician always has one hand in his pocket.'

Dry hand-to-hand resistance can be as high as 100 kΩ depending on how thick the skin is, but this can drop to 1 kΩ for wet or broken skin. This means that voltages up to 60 V are safe. The problem with high voltages is the breakdown voltage of the skin itself. The outer layer of the skin, the stratum corneum, breaks down at about 450–600 V, leaving only about 500 Ω hand-to-hand resistance.

Another problem with high voltages is the amount of energy stored by capacitances at high voltages. Energy stored increases with the square of the voltage. Consider a 1 µF capacitor charged to 30 kV: the energy stored is $\frac{1}{2}CV^2 = 0.5 \times 1$ µF $\times 30$ kV $= 450$ J. This is a large amount of energy. Medical defibrillators only use a few hundred joules, so this is more than enough to stop the heart.

## 5.3 Earthing systems

High-voltage systems are made safe by connecting all live electrodes to ground. This is achieved using multiple earthing systems: interlocked earthing arms that are automatically applied; earth switches that can be locked in place with an interlock key; and/or earth sticks that are applied by the operator.

## 5.4 Interlock systems

Interlock systems are in place to protect both personnel and the equipment. They are a combination of mechanical locks, switches and relays. Interlocks force personnel to operate the system in a specific order. Operation is prevented until all necessary actions have been carried out. Access is prevented until the system has been made safe.

## 5.5 Other hazards related to high voltage

X-rays can be produced by bremsstrahlung. Any electrons produced in a vacuum vessel can be accelerated by the applied high voltage and generate X-rays when they hit the anode. X-rays can be produced with voltages as low as 20 kV. In fact, bremsstrahlung does not have a low-voltage cut off – it is just that, below 20 kV, the configuration is usually shielded by the vacuum vessel walls. At higher voltages, shielding may become necessary, so X-ray detectors should be used during system commissioning above 20 kV.

Electric sparks can ignite hydrogen and cause an explosion. European ATEX directives [1] must be complied with.

High-voltage capacitors that are not being used should have their terminals shorted together. This is because the dielectric inside can have some 'memory' of the previous energization. Previously used high-voltage capacitors just sitting on a shelf can charge up enough to give an electric shock. The same is true for long lengths of high-voltage cable, which can have a significant capacitance.

## 5.6 HV safety design rules

The most fallible part of any system is the person operating it. High-voltage safety systems must be designed to make them idiot-proof. For regular use it is not acceptable to have safety reliant on an operator correctly following a procedure. Complacency can be a problem; the operator might have started and shut down the system thousands of times. It is important that the system is designed in such a way that it makes it impossible for an absent-minded operator to hurt themselves or others. These design rules must be followed:

### 5.6.1   *Impossible to accidentally lock someone in the HV area*

If the HV area is very small or if the high-voltage chassis (section 4.6.2) design approach is used, then this rule takes care of itself. However, if the HV area is large enough that someone could be hidden behind equipment, then a 'Search' button must be implemented. The interlock system makes it impossible to shut the door to the HV area until the 'Search' button has been pressed. The 'Search' button should be positioned at the far end of the HV area. This forces the operator to go to the far end of the HV area where they will see if anybody is in the area. Simply calling out 'Is there anybody in here?' is not acceptable because someone could be unconscious.

It must be possible to open the door from inside the HV area, even if it has been locked from the outside. This is usually implemented using the same technology as for fire escapes: crash-bars or break-glass latches. Exit signs must be provided.

### 5.6.2   *Ability to shut down the power inside and outside the HV area*

In case of emergency it should be possible to shut down the power to the high-voltage system with an emergency shut-off switch. These are usually red push buttons that lock when pushed, requiring a key to reset them. They should be positioned inside and outside the HV area and be clearly labelled.

### 5.6.3   *Impossible to power on the HV without locking the area*

The HV power supplies should have some kind of interlock that prevents them being energized until the doors to the HV area have been secured. This can be done with interlock switches or key exchanges; often it is best to use both. It should be impossible to remove the earth connection to the HV platform until the HV area has been secured.

### 5.6.4   *Impossible to enter the HV area without making it safe*

The interlock switches and key exchanges should make it impossible to open the door with the high-voltage power supplies on. However, the equipment could still be charged to a high voltage. When the door to the HV area opens, an automatic earthing arm should short the high-voltage platform to ground. If there are no capacitances or stored energy in the system, then an automatic earthing arm may not be necessary. This should be thoroughly risk-assessed.

## 5.7   Reasonably practicable

Of course, like any health and safety system, the safety design rules covered in the previous section can only be implemented as far as is reasonably practicable. Safety systems should be designed to prevent people unfamiliar with the equipment or absent-minded people from hurting themselves. You can never stop wilful people from bypassing interlocks or climbing over fences. Make sure there are signs and warning messages on all equipment, doors and entrances. That way, if anybody does bypass the safety systems, they will have only themselves to blame.

## Acknowledgements

I would like to thank David Findlay and Alan Letchford for their excellent proof-reading skills and suggestions for improvement; and Mike Perkins and Scott Lawrie for their supportive comments.